\def\gtorder{\mathrel{\raise.3ex\hbox{$>$}\mkern-14mu
    \lower0.6ex\hbox{$\sim$}}}
\def\ltorder{\mathrel{\raise.3ex\hbox{$<$}\mkern-14mu
    \lower0.6ex\hbox{$\sim$}}}

\documentclass[twocolumn]{aastex631}
\usepackage{graphicx}
\usepackage{blindtext} 
\usepackage{amsmath}
\usepackage{ulem}
\usepackage{multirow}

\def\gtorder{\mathrel{\raise.3ex\hbox{$>$}\mkern-14mu
    \lower0.6ex\hbox{$\sim$}}}
\def\ltorder{\mathrel{\raise.3ex\hbox{$<$}\mkern-14mu
    \lower0.6ex\hbox{$\sim$}}}

\begin{document}

\title{{Modeling Cosmological Evolution of Jetted Seyfert Galaxies for $z\ltorder 10$}}

\author[0009-0003-4860-8488]{Julianne Goddard}
\affiliation{Department of Physics and Astronomy, University of Kentucky, Lexington KY 40506-0055, USA}

\author[0000-0002-1233-445X]{Isaac Shlosman}
\affiliation{Department of Physics and Astronomy, University of Kentucky, Lexington KY 40506-0055, USA}

\author[0000-0002-0071-3217]{Emilio Romano-Diaz}
\affiliation{Argelander-Institut f\"ur Astronomie, Auf dem H\"ugel 71, 53121 Bonn, Germany}

\begin{abstract}
We use high-resolution cosmological zoom-in simulations to model mechanical and thermal feedback from AGN onto the evolution of Seyfert-type galaxies, studying the morphology of central galaxies growing within dark matter (DM) halos with masses log\,$M/M_\odot\sim11.8$ at $z=0$. In Paper\,I, we focused on the end products at $z=0$, here we analyze evolution for $z\ltorder10$. Black holes (SMBHs) of $\sim10^6\,M_\odot$ were seeded at $z\sim9.1$ and $z\sim3.7$, producing jets along their spin axes. Obtained SMBH accretion rates vary in the range $\sim0.3-10^{-4}$ of the Eddington rate. We compared the basic properties of galaxies, such as star formation rate, masses, gas and stellar fractions, bulge-to-disk mass ratios, SMBH masses, etc., over the range of redshifts.  Our results indicate that jets and associated over-pressured bubbles have substantial effects on Seyfert galaxy evolution, including properties of the interstellar and circumgalactic medium (ISM and CGM), and even beyond.  This feedback can suppress and even quench star formation, reduce stellar mass and gas fraction, modify the bulge-to-disk ratios, drive outflows from galaxies and host DM halos, and metal-enrich the CGM. The jets are largely but not exclusively contained within galaxies. However, over-pressured bubbles cross and modify the composition of the CGM and IGM, their thermodynamic and dynamic state, and generate vorticity. The CGM emerges as a complex region, where action of galactic outflows and jet-formed bubbles combines with the influx from cosmological filaments and diffuse accretion. Ultimately, the above processes affect the gas balance within the galaxy, its morphology,  and gas supply to the SMBH, limiting its growth.
\end{abstract}


\section{Introduction}
\label{sec:intro}

Supermassive black holes (SMBHs) permeate the universe and can be found in most massive galaxies. Those that are accreting, are observed as active galactic nuclei (AGN). Active SMBHs have been detected up to $z\sim 10$ \citep[e.g.,][]{wu15,venemans17,banados18,bosman23,bogdan23,larson23,maiolino24}, and their origin includes a narrowing range of options \citep[e.g.,][]{begelman09,inayoshi20,volonteri21}. Recent high-$z$ studies have revealed that SMBHs grow more quickly than anticipated and bring to light new questions about how and when SMBHs have been seeded \citep[e.g.,][]{begelman06}.  

A number of scaling relations have been found to relate the SMBH properties to those of their host galaxies. These relationships are based on solid estimates that the SMBHs release substantial amounts of energy, radiative and mechanical, which can compete with the binding energy of the baryonic components in galaxies. However, how exactly this energy is digested by baryons is still far from being understood. Strong indications exist that AGN feedback affects the high--mass end of galaxies, possibly in tandem with curtailing the growth of SMBHs themselves. However, for less massive SMBHs, residing in Seyfert galaxies, the effect is studied much less.  

The focus of this work is (1) to quantify the effects of the mechanical and thermal feedback from Seyfert jets onto the host galaxy morphology, both stellar and gas, for $z\ltorder 10$, and (2) to analyze the effect of SMBH seeding time on host galaxies. Relativistic jets have traditionally been associated only with the highest mass SMBHs and their high-luminosity outflows. However recent observations have identified an ever-increasing number of jets emitted from lower-luminosity, $< 10^{46}\,{\rm erg\,s^{-1}}$ Seyfert-type AGN.  These are identified in radio, as either extended or compact emission, which cannot be accounted for by the star formation (SF) alone \citep[e.g.,][]{komossa06,caccianiga15}, or detected from 37\,Ghz emission to $\gamma$-rays \citep[e.g.,][]{abdo09,lahteenmaki18,romano18,rakshit21b}.  Jets have been identified predominantly in narrow-line Seyfert\,1 (NLS1) galaxies, characterized by low mass, $M_\bullet < 10^8\,M_\odot$, SMBHs and high Eddington fraction accretion, $f_{\rm Edd} > 0.1$ \citep[e.g.,][]{varglund22}. Few cases of relativistic jets were identified in type-2 or intermediate-type Seyfert galaxies \citep{jarvela20,Foschini2020}. Compact jets are also observed in compact steep spectrum (CSS) and Ghz-peaked spectrum (GPS) radio sources, many of which have $M_\bullet < 10^8\,M_\odot$ \citep[e.g.,][]{nascimento22,marques24}. 

AGN can interact with their host galaxies exerting a variety of feedback, by radiation, collimated outflows (jets) and accretion disk winds \citep[][and refs. therein]{fabian12}.  Disk winds can be driven radiatively \citep[e.g.,][]{shlosman85,murray95,arav97}, thermally \citep[e.g.,][]{begelman83}, by cosmic rays \citep[e.g.,][]{begelman91}, hydromagnetically \citep[e.g.,][]{blandford82,emmering92,contopoulos95,bottorff97}, or by a combination of these factors. Moreover, the MHD winds provide a natural explanation for the {\it outflowing} `torii' in AGN \citep{elitzur06}, in addition to the extraction of angular momentum from the disk.

It is plausible that this energy deposition in the host AGN galaxy leads to the observed scaling relations at low redshifts, but the details are elusive. Equally unclear are the timescales under which these relations are being established,  it could be as long as few Gyrs or even longer. For example, the AGN can in principle expel the gas from the galactic bulge, terminating the SF there and cutting off the accretion flow towards the SMBH. But is this process recurrent and how long does it take to renew the bulge growth (if at all)? On the other hand, if the gas replenishing process proceeds from outside in, what prevents the bulge from resuming its growth while the SMBH is dormant? An alternative view is that this correlation has been established as a result of the merger activity \citep[e.g.,][]{jahnke11}. 

A closely related issue is the extent of mechanical feedback from AGN in Seyfert galaxies. More powerful jets in quasars break out of the host galaxies to typical distances of $\sim $Mpc. Typical sizes of {\it observed} jets in nearby Seyferts are of kpc-scale, but these jets are oriented into the galactic disks and hence suffer from maximal resistance and dissipation. Observations of jets with different orientations become more frequent \citep[][Paper\,I, and refs. therein]{goddard25}. Moreover, jets in local Seyferts have been observed to interact strongly with their hosts. They can shock surrounding gas launching large scale outflows and form over-pressured bubbles around galaxies \citep[e.g.,][]{tadhunter14,marques24}.  These outflows and bubbles can have a positive or negative effect on SF \citep[e.g.,][]{krause23}, influence the interstellar medium (ISM), gas morphology \citep{Nesvadba2021,nandi23}, impact the thermodynamic properties of ISM and the circumgalactic medium (CGM) (Paper\,I), and affect the distribution of metals in and around the host \citep[e.g.,][]{cunlow04}.   

However, until recently, space- and ground-based telescopes have lacked resolution to identify Seyferts further than $z\sim 0.8$ directly.  The latest instruments, especially JWST, are providing the first H$\alpha$ spectroscopic studies of moderate luminosity AGN hosts at $z\sim 4-7$ \citep[e.g.,][]{pacucci23,maiolino24}, which also violate the $M_*-M_\bullet$ relation. Although resolution is still limited, we can in many cases obtain rough estimates for morphology, mass, SF rates (SFR), SMBH masses $M_\bullet$, and the mechanical luminosity of jets, $L_{\rm jet}$. 

Many of the traditional methods of jet identification in local Seyferts are, as yet, unavailable at high redshift. New strategies are being developed to circumvent these limitations, e.g., \citet{rakshit21} use correlation between H$\beta$ and Mg\,II lines to identify NLS1’s at $z > 0.8$ in the SDSS data, where H$\beta$ could not be detected. Perhaps, the best way to identify small-scale jets at high $z$ is to find the NLS1s and search for associated $\gamma$-rays.

A very limited number of studies about relatively high-$z$ jetted Seyferts are known.  A collection of suspected NLS1 galaxies for $z > 0.8$ have been observed with extended FR\,II-type radio emission, while local NLS1s have a higher Eddington fraction and are more radio quiet, with FR\,I type radio emission \citep{rakshit21}. \citet{wolf23} similarly observed a suspected NLS1 at $z > 6$. However, these recently observed galaxies are already of similar stellar and SMBH masses to typical Seyferts at $z=0$, and thus likely represent the progenitors of a more massive population of galaxies in the local universe. The precursors of local jetted Seyfert galaxies are unlikely to be found observationally without a theoretical basis for their identification.  

Numerical simulations of jets have reproduced some of the observed phenomena. \citet{kawata05} explained observed X-ray/optical properties of elliptical galaxies by injecting thermal energy in the galaxy centers at $z<1$. Motivated by semi-analytic models, \citet{okamoto08} implied radiatively-inefficient AGN mode to produce a strong feedback in galaxies with low SFRs, using a subgrid recipe for the energy injection. \citet{irodotou22} modeled the effect of the `radio' mode feedback using expanding bubbles, comparing it with an isotropic wind, finding that increasing feedback decreases stellar mass and alters galactic morphology. \citet{byrne24} applied jet feedback in $10^{12}-10^{13}\,M_{\odot}$ halos confirming a negative impact on the SF and trending towards a more spheroidal distribution. \citet{appleby21} analyzed the effects of jet feedback on the CGM, detecting a reduced baryon fraction and increase in metallicity. \citet{goddard25} presented cosmological simulations of Seyferts at $z=0$, subject to jet activity and expanding bubbles, and their effect on the SFRs, stellar masses, metallicities, etc., as a function of the feedback strength.  By simulating jets over relatively short times, $\ltorder 1$\,Gyr, and focusing on the jet-driven outflows, \citet{mukherjee18} observed that the jet angle and strength can affect the SFR, potentially producing both positive and negative feedback.  This was confirmed by \citet{talbot22}, who studied outflow effects on the CGM by low-power jets, finding a strong influence of the jet direction on the outflow. 

The goal of this work is to model the mechanical and thermal feedback from jetted Seyfert galaxies from $z\ltorder 10$ to $z=0$. We aim at analyzing the effect of jets on the host galaxy morphology over cosmological times. In Paper\,I, we analyzed the results of high-resolution cosmological zoom-in simulations at $z=0$. Here, using the same simulations, we focus on redshift evolution of these objects by varying AGN and SN feedback and the seeding time of the SMBHs. 

This paper is structured as follows. Section\,2 describes the numerics and initial conditions for our modeling, section\,3 provides the results, which are summarized, discussed and compared to observations in section\,4. 

\section{Numerics and simulation setup} 
\label{sec:numerics}

We have performed a suite of high-resolution cosmological zoom-in simulations employing the $N$-body/hydro code \textsc{gizmo}, using the MFM hydro solver \citep{hopkins15}. The initial conditions (ICs) of the parent simulation used the \citet{planck16} $\Lambda$CDM concordant model, with $\Omega_{\textrm m} = 0.308$, $\Omega_\Lambda = 0.692$, $\Omega_{\textrm b} = 0.048$, $\sigma_8 = 0.82$, and $n_{\textrm s} = 0.97$, with the Hubble constant $h = 0.678$ in units of $100\,{\textrm {km}\,\textrm s^{-1}\,\textrm {Mpc}^{-1}}$. The ICs were generated at $z=99$ using the \textsc{music} code \citep{hahn11} within a box of $50\,h^{-1}$\,Mpc, and were evolved to $z=0$.

From the parent, uni-grid, DM-only simulation, a halo has been chosen for re-simulation at a higher resolution. This halo has been identified by the group finder \textsc{rockstar} \citep{behroozi12}, with a Friends-of-Friends (\textsc{FoF}) linking length of $b=0.28$. The DM-{\it only} halo virial radius, $R_{\textrm {vir}}$, and its virial mass, $M_{\textrm {vir}}$, have been defined in terms of $R_{200}$ and $M_{200}$ \citep[e.g.,][]{navarro96}, where $R_{200}$ is the radius within which the mean interior density is 200 times the critical density of the universe at that time, and $M_{200}$ is the corresponding enclosed mass. A DM halo with log\,$M_{\rm vir}/M_\odot = 11.8$, $R_{\rm vir} =231$\,kpc, halo spin $\lambda = 0.03$ and a local environment with a relative overdensity $\delta = 2.0$ has been selected at $z=0$. 

The zoom-in ICs are composed by five nested refinement levels on top of the base grid, i.e., from $2^7$ to $2^{12}$. The DM-only version was first evolved in order to check for, and avoid contamination from massive, lower-resolution particles in the highest resolution-level volume. The baryons were included at the highest level of refinement in the reconstruction of their ICs.  

The effective number of particles (DM and baryons) in our simulations is $2\times 4,096^3$, resulting in the mass resolution per particle of $3.6\times 10^4\,{\rm M_\odot}$ for gas and stars, and $1.9\times 10^5\,{\rm M_\odot}$ for the DM. The minimal adaptive gravitational softening in comoving coordinates for the gas is 1\,pc, for stars 20\,pc and 200\,pc for DM. 

Galaxies have been identified by the group-finding algorithm \textsc{hop} \citep{eisenstein98}, using the outer boundary threshold of baryonic density of $10^{-4}\,n^{\textrm {SF}}_{\textrm {crit}} = 10^{-2}\,{\rm cm^{-3}}$, which ensured that both the host SF gas and the lower density non-SF gas are roughly bound to the galaxy \citep{romano-diaz14}. This assures that identified galaxies are not imposed with a particular geometry. Following Paper\,I, the \textsc{hop}-defined galaxy size was found to be similar to 0.1$R_{\rm vir}$, which is used here.

Gas heating and cooling from $10^{10}$\,K down to 10\,K are implemented, including H and He ionization$+$recombination, collisional, free-free, dust collisional, cosmic ray, and Compton effects, as well as metal-line \citep{wiersma09}, fine-structure, and molecular cooling \citep{hopkins18,hopkins22}. Metal enrichment is included: the metallicity increases in the SF gas and scales with the fraction of stars that turn into SN, and the metal yield per SN (see below). A total of 11 metal species were followed in gas and stars, including H, He, C, N, O, Ne, Mg, Si, S, Ca, and Fe. The H$_2$ abundances used for cooling calculations are estimated from the \citet{krumholz11} analytic fitting function. Metal diffusion is not implemented explicitly, but metals can be transported by mechanical feedback from SN and AGN (see below). Our simulations include the redshift-dependent cosmic UV background \citep[e.g.,][]{faucher-giguere20,shen20}. 

The SMBHs have been modeled with sink particles that were seeded once the stellar mass of the galaxy has reached the target value of $10^{8}\,M_\odot$, or alternatively when the DM halo mass has reached $10^{11}\,M_\odot$. These models have been abbreviated as the early seeded SMBH (EBH) at $z\sim 9.1$, and the late seeded SMBH (LBH) at $z\sim 3.7$, respectively, and allow us to access the effect of a different seeding redshift. We include an artificial velocity-damping term to continuously move the SMBH toward the most bound particle \citep{wellons23}. The SMBHs were seeded with $M_\bullet\simeq10^6\,M_\odot$, inherit the angular momentum from accreted material, and grow by accreting surrounding gas, with an accretion rate, $\dot M_{\rm grav}$, based on gravitational torques \citep{shlosman89}, and calculated using \citet{hopkins11} method. The spin axis of the SMBH is determined by the cumulative angular momentum vector of the accreting matter, which is followed by the code.

We have modified the SMBH accretion rate prescription using the efficiency parameter $\epsilon$ from \citet{angles-alcazar17}, to make $\dot M_{\rm grav}$ more realistic and to capture the effects of unresolved processes affecting gas inflow that are not addressed in the existing subgrid recipe. For each seeded SMBH, the sole parameter varied between the AGN models was $\epsilon$, set to $\epsilon=0$, 5\%, 15\%, and 50\%. These models have been denoted as $\epsilon_0$, $\epsilon_5$, $\epsilon_{15}$ and $\epsilon_{50}$, respectively (see Table\,\ref{tab:models} for model abbreviations).  So the final accretion rate is
\begin{equation}
    \dot{M}_\bullet=\epsilon\dot{M}_{\rm grav}.    
\end{equation}
The SMBH accretion rate measured is as a fraction of the Eddington accretion rate, $\dot M_{\rm Edd}$, is 
\begin{equation}
    f_{\rm Edd}=\dot M_\bullet/\dot M_{\rm Edd},
\end{equation}
where $\dot M_{\rm Edd}=L_{\rm Edd}/c^2$, and $L_{\rm Edd}$ is the Eddington limit for $M_\bullet$. 

To model the AGN jets, we use the hyper-refined particle spawning \citep{torrey20}, modified by \citet{su21}. Particles of mass $10^3\,M_\odot$ are spawned at a fixed fraction of the SMBH kernel radius with an initial velocity of $3\times 10^4\,{\rm km\,s^{-1}}$ and a temperature of $10^{10}$\,K. Our choice of jet particle mass follows from numerical considerations, while the velocities have been chosen so that the jet energies lie within the observed range of Seyfert galaxies. The initial temperature has been taken to keep the jet particle energy dominated by its kinetic energy, but also with non-negligible thermal energy. However, independent of initial temperature, the jet particles will immediately shock on the ISM and heat up to about $10^{10}$\,K. These particles are launched along the spin axis of the SMBH with a mass loading $\eta$=0.1, meaning 10\% of the accreted gas is returned in the form of jet particles.  The launching rate is thus $\dot{M}_{\rm jet}=\eta\dot{M}_\bullet$, giving a jet energy injection rate of 
\begin{equation}
    L_{\rm jet}=\eta[1/2 \dot M_\bullet v_{\rm jet}^2 +3/2 (\dot M_\bullet/m_{\rm p}) kT],
\label{eq:Ljet}
\end{equation}
where $v_{\rm jet}$ is the jet particle velocity, and $T$ is their temperature. Eq.\,\ref{eq:Ljet} describes both kinetic and internal energies of the jet particles. The velocities of the spawned particles are initially perfectly collimated. We refer to this feedback as a 'jet' feedback. Once created, the jet particles interact hydrodynamically in the same way as any of the other gas particles. When they decelerate to at least one fourth of their initial velocity, and enter the kernel radius of another gas particle in a head-on trajectory, the jet particles re-merge and the mass-weighted properties of the two particles are averaged.  The jet particles inherit their metallicity from gas particles closest to the SMBH when they spawn. 

The density threshold for star formation (SF) was set to $n^{\textrm {SF}}_{\textrm {crit}} = 100\,{\textrm {cm}^{-3}}$. We use the \citet{chabrier03} Initial Mass Function (IMF) and the mechanical feedback from SN--type\,II, as given in \citet[][see also Paper\,I]{hopkins18}. In LBH models, each SN event injects energy of $1\times 10^{51}$\,erg, mass of 14.8\,$M_\odot$, and a metal mass of 2.6\,$M_\odot$ into the surrounding gas within a radius of 200\,pc. For EBH models, each SN event injects energy of $5\times 10^{51}$\,erg over 10\,Myr within 700\,pc, mass and metal injection remains unchanged. The motivation for adjusting the SN feedback prescription in the EBH models comes from previous studies \citep[e.g.,][]{roca21} that show similar parameters are able to replicate observations well.  This change to the stellar feedback routine introduces another parameter to the comparison between our EBH and LBH models. Therefore, we have run an additional test model which uses the original stellar feedback prescription from the LBH models, but seeds the SMBH using the EBH model criteria. This test model, $\epsilon_{\rm 5, EBH}$(test), is listed in Table\,\ref{tab:models} and discussed in detail in the Results section.

Additional technical details regarding the simulation setup, the star formation and supernovae (SN) feedback, as well as the seeding of the SMBH and its mechanical and thermal feedback in the form of jets, have been outlined in Paper\,I. Hot over-pressured bubbles form as a result of jet interaction with the ambient gas --- these were called `cocoons' in Paper\,I.

\begin{deluxetable}{ccccccc}[ht!]
\tablecolumns{4}
\tablecaption{List of Models}
\tablehead{
\colhead{} & \colhead{Feedback Type} & \colhead{ $\epsilon_0$} & \colhead{$\epsilon_5$} & $\epsilon_5$(test) & $\epsilon_{15}$ & $\epsilon_{50}$ }
\startdata
\multirow{3}{2em}{\rotatebox{90}{\bf EBH}} & {AGN}   & 0   & 0.05 & 0.05 & 0.15 & 0.5  \\
   & {SN}           & 5   & 5    & 1    &  5   &  5   \\
   & BH seeding ($z$) & --- & 9.1  & 9.1  & 9.1  & 9.1  \\
\hline
\multirow{3}{2em}{\rotatebox{90}{\bf LBH}}  & {AGN}             & 0   & 0.05 & ---  & 0.15 & 0.5 \\
  & {SN}               & 1   & 1    & ---  & 1    & 1   \\
  & BH seeding ($z$)   & --- & 3.7  & ---  & 3.7  & 3.7 \\
\enddata
\tablecomments{AGN feedback is given by $\epsilon$ defined in section\,\ref{sec:numerics}; SN feedback is given in units of $10^{51}\,{\rm erg\,s^{-1}}$; Seeding time of the SMBH is at redshift $z$. In the text, we abbreviate each model by a subscript EBH, i.e., early BH, or LBH, i.e., late BH, depending on the seeding time.}
\label{tab:models}
\end{deluxetable}

\section{Results}
\label{sec:results}

In this section, we provide the results of our zoom-in cosmological simulations and present the observed effects of mechanical and thermal feedback from AGN jets and the associated expanding bubbles onto the evolution of their Seyfert-type host galaxies over cosmological timescales, from $z\sim 10$ to $z=0$. We have carried out two suites of simulations, seeding an SMBH when the host halo mass reaches $\sim 10^{11}\,M_\odot$ at $z\sim 3.7$, and only varying the efficiency of the SMBH accretion, $\epsilon$ --- these models have been named LBH. The second suite evolves the same galaxy, but exploring an earlier SMBH seeding, when the stellar mass of the host reached $\sim 10^{8}\,M_\odot$ at $z\sim 9.1$, and with a stronger SN feedback --- named EBH. 

To emphasize our goal of comparing evolution of the AGN and non-AGN models, we normalize various parameters of the AGN models in units of the non-AGN (or no-SMBH) models. Such normalized models must be approached with caution --- the normalized evolution should be only analyzed and interpreted keeping in mind that the actual, non-normalized evolution may differ profoundly. Specifically, we compare the evolution of $\epsilon_5$, $\epsilon_{15}$ and $\epsilon_{50}$ models with $\epsilon_0$ models in both EBH and LBH sequences. 
 
\subsection{Host DM Halo and galactic stellar mass evolution}
\label{sec:evolution}

\begin{figure}[ht!]
\begin{center}
\includegraphics[width=.9\linewidth]{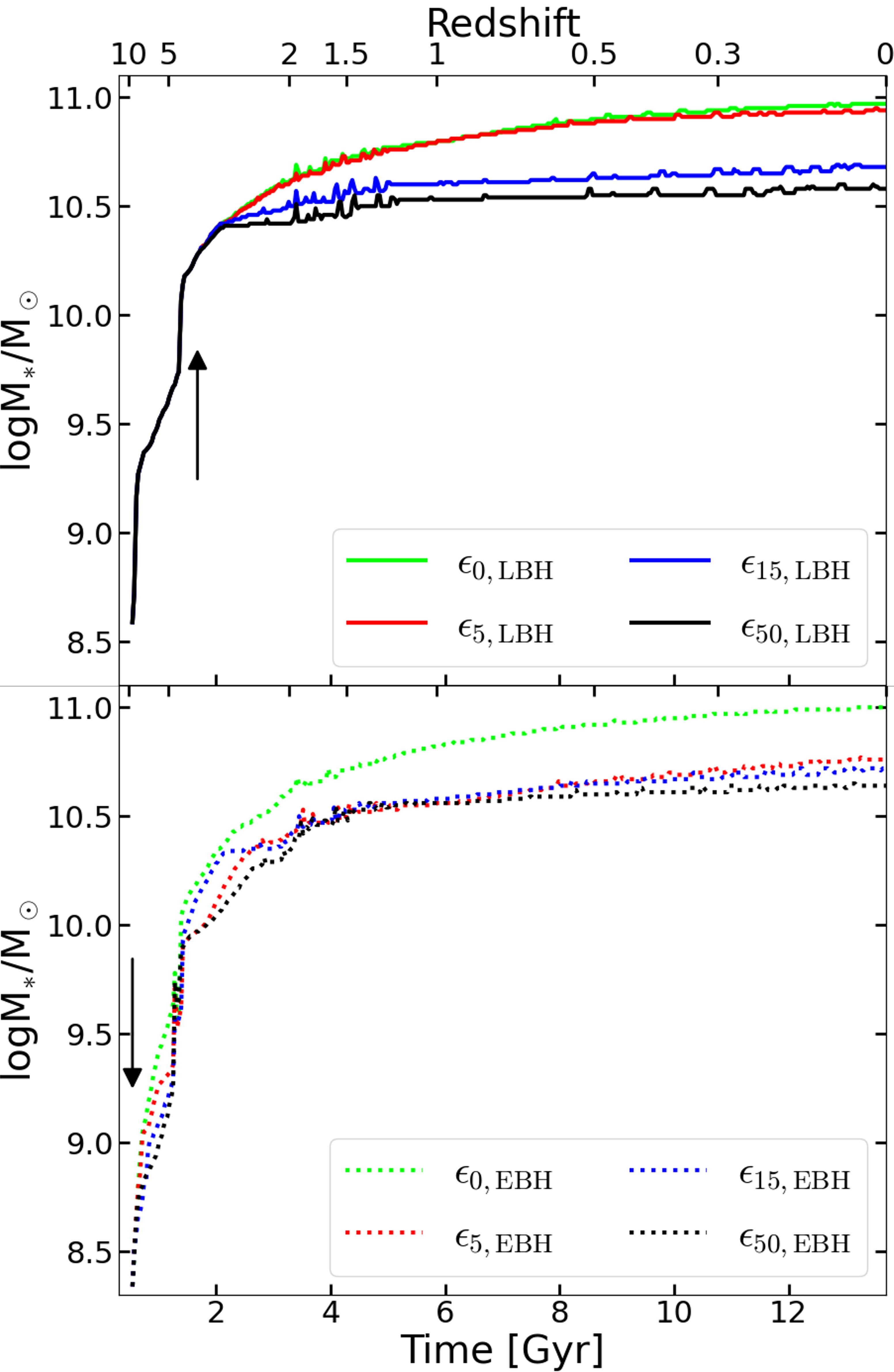}
\end{center}
\caption{Evolution of stellar mass, $M_*$ on a galactic scale, $\sim 0.1R_{\rm vir}$, for LBH (top frame) and EBH (bottom frame). The black arrows indicate the seeding time of the SMBHs.
\label{fig:starmass}}
\end{figure}

$M_{\rm vir}$ and $R_{\rm vir}$ evolution are not affected by the presence of the AGN or its seeding time, to any detectable level. The halo forms at $z\sim 2$, when it reaches log\,$M_{\rm vir} \sim 11.5\,M_\odot$, half of its value at $z=0$. This redshift appears to divide also the evolution of the central galaxy into substantially different stages --- shown later in this section and discussed in section\,\ref{sec:discuss}. Only around $z\sim 4$ do the parent halos experience a major merger. Intermediate and minor mergers are more frequent.  
 
Evolution of the stellar mass of the central galaxy, $M_*$, is strongly affected by the presence of AGN (Fig.\,\ref{fig:starmass}).  Both the early-seeded (EBH) and late-seeded (LBH) models display a substantial decrease in the stellar mass growth due to the presence of AGN.  In all cases, $M_*$ inversely correlates with the AGN feedback, $\epsilon$ --- its increase leads to a lower final stellar mass.  Galactic $M_*$ starts to diverge immediately with the seeding time from $\epsilon_0$ models. By $z=0$, the $\epsilon_0$ models have $\sim 2.5$ times the stellar mass compared to the AGN models. When comparing between the EBH and LBH models, the latter appear only slightly less massive at the end of the evolution. The model $\epsilon_5$, behaves differently in EBH and LBH runs, and we return to this issue in the later sections. These differences in the final galaxy stellar properties follow from the SF history which in turn is expected to correlate with the gas properties in galaxies (see section\,\ref{sec:SF}). 

\begin{figure}[ht!]
\center
\includegraphics[width=0.9\linewidth]{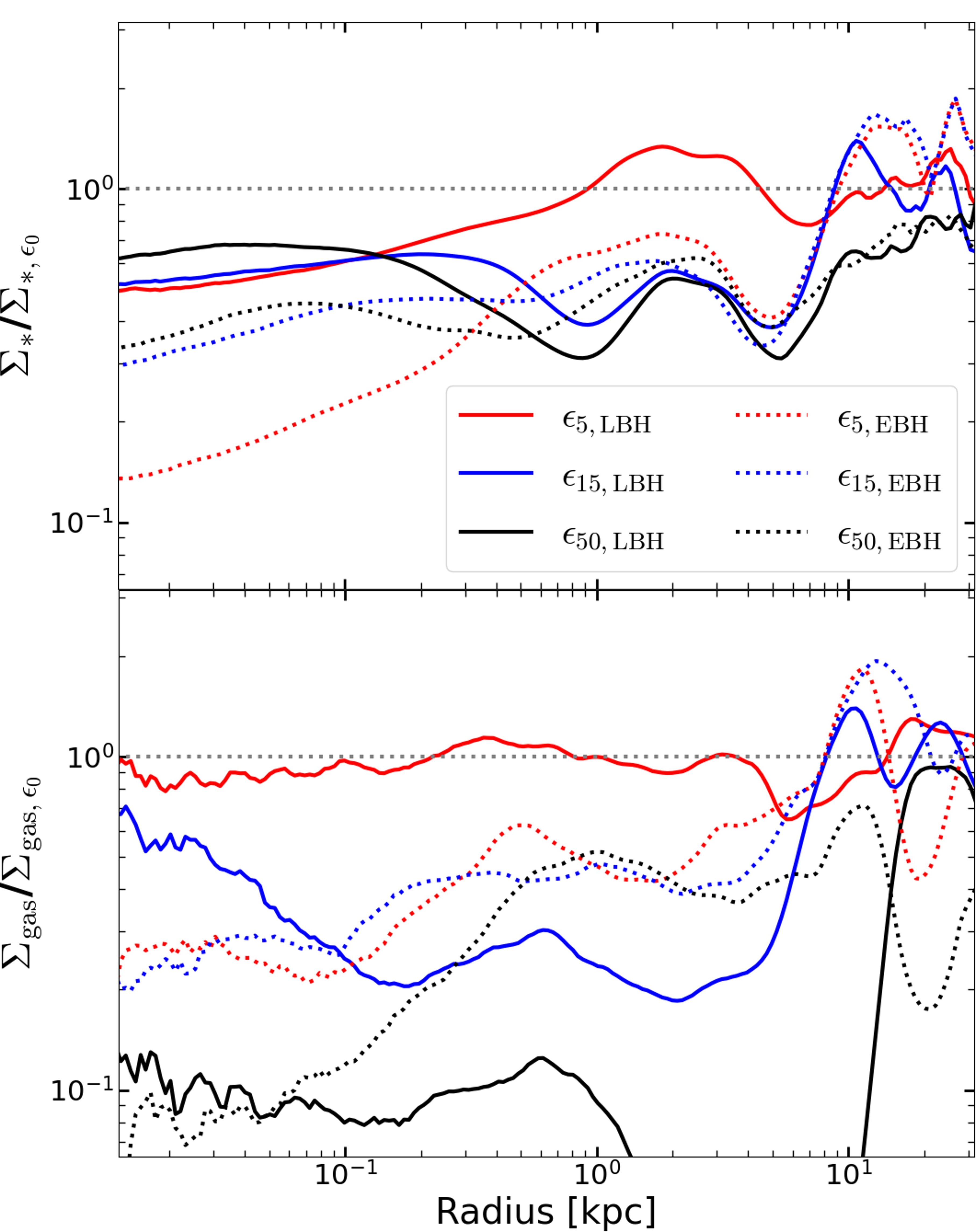}
\caption{Surface density distribution for stars, $\Sigma_*(r)$ (top), and gas, $\Sigma_{\rm gas}(r)$ (bottom), within the galaxy, time-averaged from $z=2$ to $z=0$ and normalized by the surface density within $\sim 0.1R_{\rm vir}$ of the $\epsilon_0$ model at each redshift. The LBH models are represented by continuous lines and EBH models by dotted lines. The horizontal dotted lines are the benchmark when the ratio equals unity. 
\label{fig:SurfDen}}
\end{figure}

In the LBH models, long-lived gaseous voids form in the galaxies as a result of the AGN feedback, and effectively quench the SF on Gyr timescales in the $\epsilon_{15}$ and $\epsilon_{50}$ models. In EBH models, these voids form in all three models, but they are shorter-lived. The reason for these gas expulsions is a combination of the SN and AGN feedbacks. The jets deposit their energy within $\sim 10-60$\,kpc, depending on the feedback strength (Paper\,I; see also discussion in section \ref{sec:SMBH}). This triggers bubble-type outflows which propagate well outside the CGM, and, for some models, on the Mpc-scale. 

In order to evaluate the time-integrated effects  of the AGN feedback on the radial distribution of stellar and gaseous masses within the modeled galaxies, we have plotted the radial profiles of surface densities, $\Sigma_*(r)$ and $\Sigma_{\rm gas}(r)$, averaged over $z=2-0$ period, i.e., when the galactic disks growth has saturated. Figure\,\ref{fig:SurfDen} displays these entities normalized by the surface density profile of the $\epsilon_0$ model of their respective set. All the AGN models, apart from the LBH $\epsilon_5$ outlier, have a reduced $\Sigma_{\rm *}$ within the inner $\sim 10$\,kpc. In this region, the $\epsilon_0$ model profiles lie above the AGN models by a factor of a few. The observed trend is that the radial stellar profile ratios rise with radius and reach unity outside this radius. The gas density profiles follow the same trend, with larger variation and larger vertical offset down, which increases with the AGN feedback. This is a clear signature of the AGN feedback.

\subsection{Star Formation History: the effect of AGN}
\label{sec:SF}

\begin{figure}[ht!]
\center
\includegraphics[width=0.9\linewidth]{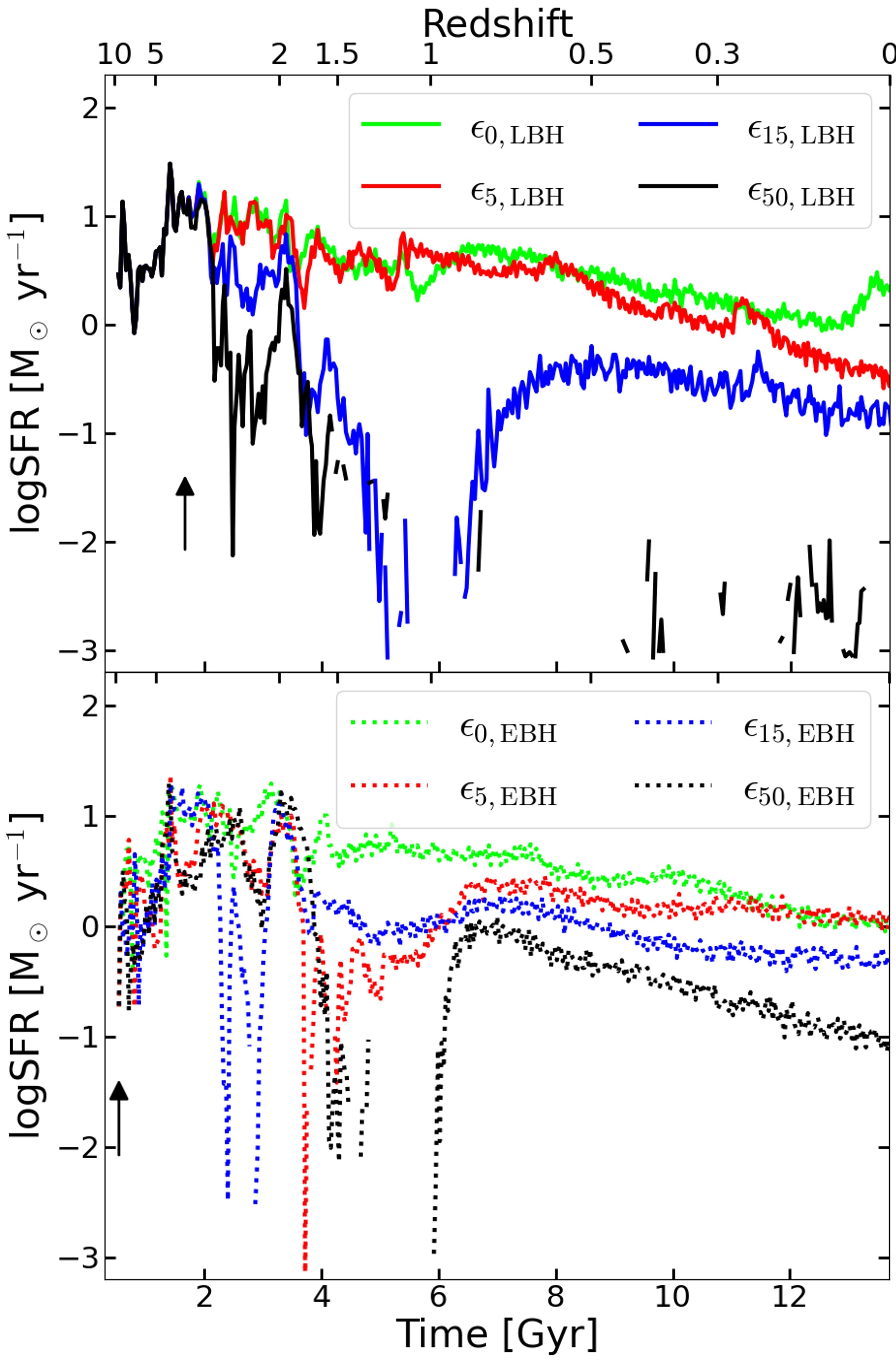}
\caption{Evolution of galactic SFR within $0.1R_{\rm vir}$ for LBH (top) and EBH (bottom) models. Each curve represents a different AGN feedback model. The black arrows indicate the seeding time of the SMBHs.
\label{fig:SFR}}
\end{figure}

\begin{figure}[ht!]
\center
\includegraphics[width=0.9\linewidth]{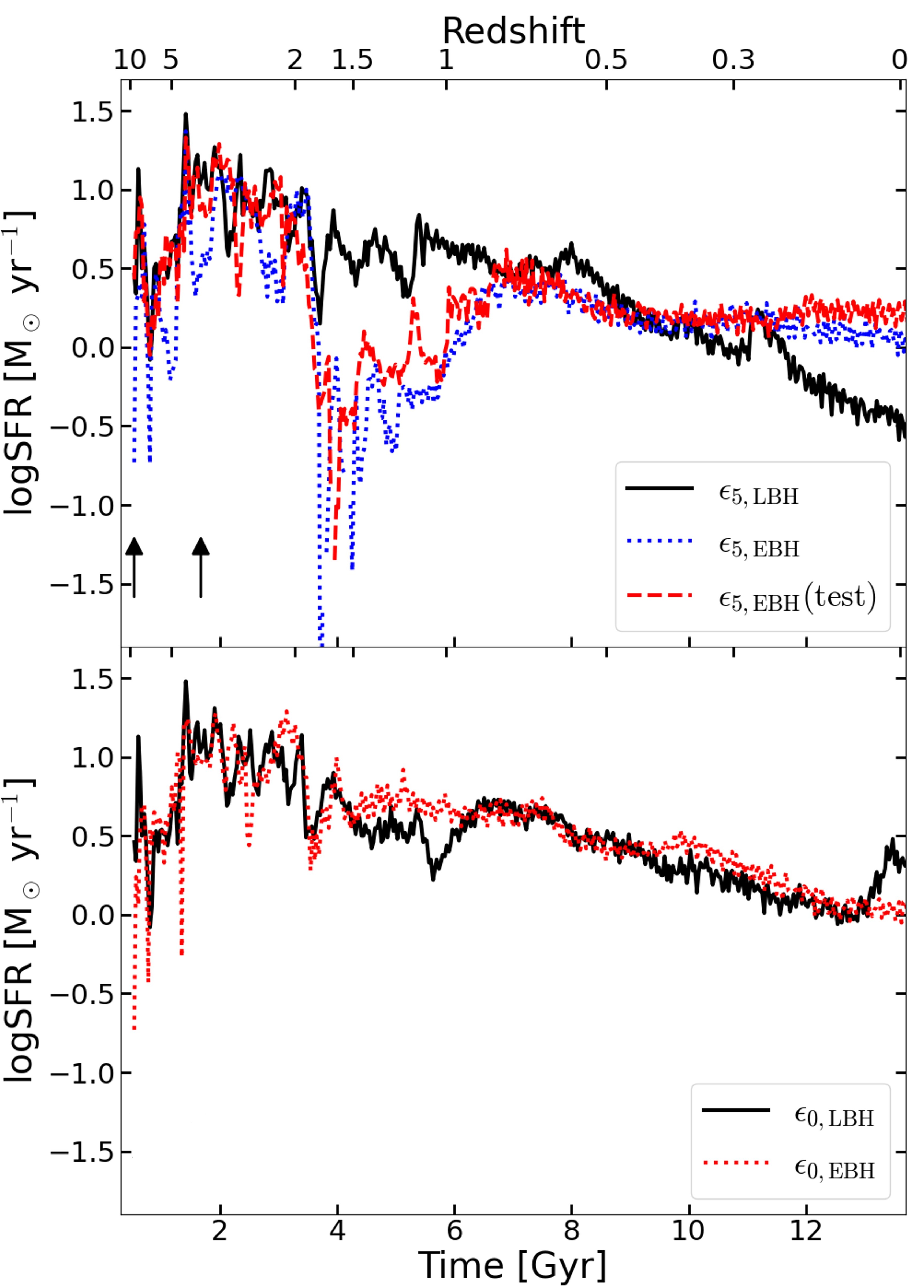}
\caption{Evolution of galactic SFR within $0.1R_{\rm vir}$ for three similar $\epsilon_5$ models (top frame) and two $\epsilon_0$ models (bottom frame). Top: The LBH model (black solid line), the EBH model with enhanced SN feedback (blue dotted line), and the comparison $\epsilon_{\rm 5,EBH}$(test) model with the same SN feedback as in LBH models (red dashed line).  The $\epsilon_{\rm 5,EBH}$ model differs from the other two by increased strength of the SN feedback, increased range of the SN feedback to 700\,pc, and a reduced 10\,Myr energy deposition timescale. Bottom: Two models without AGN. The EBH model has enhanced SN feedback (see section\,\ref{sec:numerics}). The black arrows indicate the seeding time of the SMBHs.
\label{fig:SFR_comp}}
\end{figure}

\begin{figure}[ht!]
\center
\includegraphics[width=0.9\linewidth]{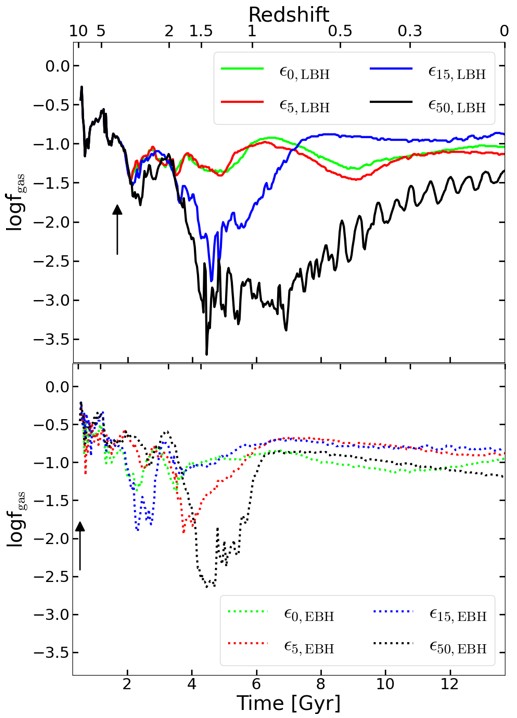}
\caption{Evolution of the gas fraction, $f_{\rm gas}$, within the galactic radii, $0.1R_{\rm vir}$, for all models, LBH (top) and EBH (bottom). The black arrows indicate the seeding time of the SMBHs. 
\label{fig:fgas}}
\end{figure}

Next, we turn to the galactic SFR in EBH and LBH models, as this is one of the most important parameters characterizing galaxy evolution (Fig.\,\ref{fig:SFR}).  In the LBH models, the SFR in $\epsilon_5$ and $\epsilon_0$ models follow a very similar evolution, except at low redshifts, where $\epsilon_5$ dives a factor of 3 below $\epsilon_0$. The reason for this can be clearly identified --- the first jet induced bubble to breach R$_{\rm vir}$ in $\epsilon_5$ has a very late appearance, around $z\sim 0.2$ (section\,\ref{sec:cocoons} and Paper\,I), when it exerts a negative feedback on the SF. The higher $\epsilon$ models display a sharp decline in SFR almost immediately with the introduction of the BH seeds, and higher $\epsilon$ models experience a substantially stronger and more prolonged decline in SFR. The $\epsilon_{\rm 15,LBH}$ and $\epsilon_{\rm 50,EBH}$ models essentially quench the SF for $\sim 2$\,Gyr, while $\epsilon_{\rm 50,LBH}$ has difficulty to recover its SFR even after 9\,Gyr.

The SFRs in the EBH models all differ from each other. Troughs in the SFR are seen in all the EBH AGN models, but they begin earlier than in the LBH models and recover more quickly. The strongest feedback, $\epsilon_{\rm 50,EBH}$ model, is quenched for $\sim 2$\,Gyr only, and displays an abrupt recovery. We observe a clear anti-correlation between the SFRs and $\epsilon$ in these models as well.  

To separate the effects of the SN and AGN feedback in our models, we have invoked two tests, which are shown in Figure\,\ref{fig:SFR_comp}. In the upper frame, we compare three $\epsilon_5$ models: two of these are the LBH and EBH models, and the third is the $\epsilon_{\rm 5,EBH}$ model that has a standard SN feedback (i.e., the SN feedback prescription used in the LBH models, see section \ref{sec:numerics}). The bottom frame compares two $\epsilon_0$ models differing in their SN feedback strength, being stronger for the EBH model, as discussed in section\,\ref{sec:numerics}.

We start with the upper frame in Figure\,\ref{fig:SFR_comp}. The $\epsilon_{\rm 5,EBH}$(test) (red dashed) and $\epsilon_{\rm 5,EBH}$ (dotted blue) models differ only in their SN feedback. For $z\gtorder 1.5$, the $\epsilon_{\rm 5,EBH}$(test) curve lies slightly above the $\epsilon_{\rm 5,EBH}$ curve, and both are bursty. Both EBH lines display a sharp decrease during $z\sim 2-0.7$ which coincides with a sequence of minor and intermediate mergers on highly inclined orbits. In tandem with SN feedback, they push the gas away from the disk plane to higher latitudes, where it  forms a disk on a polar orbit and becomes a target to the jet and expanding bubbles --- this affects and typically reduces the SFR. At lower redshifts, these curves in Figure\,\ref{fig:SFR_comp} nearly merge. The final $M_*$ in these models differ by $\sim 20\%$, due to the small differences between these models. 

The $\epsilon_{\rm 5,LBH}$ model does not follow the other two models in the upper frame of Figure\,\ref{fig:SFR_comp} and exhibits no SFR trough around $z\sim 2-0.7$. Moreover, it dives below the other two models for $z\ltorder 0.3$, by a factor of $\sim 4$. As $\epsilon_{\rm 5,LBH}$ and $\epsilon_{\rm 5,EBH}$(test) models have identical SN feedback, different evolutionary tracks, which are separated up to a factor of $\sim 30$ in the trough region, must be attributed solely to the AGN feedback, and specifically to the seeding time of the SMBH. Therefore, we conclude that {\it varying} the SN feedback prescription in our modeling has only a minor effect on evolution of the SFR. 

\begin{figure}[ht!]
\center
\includegraphics[width=0.9\linewidth]{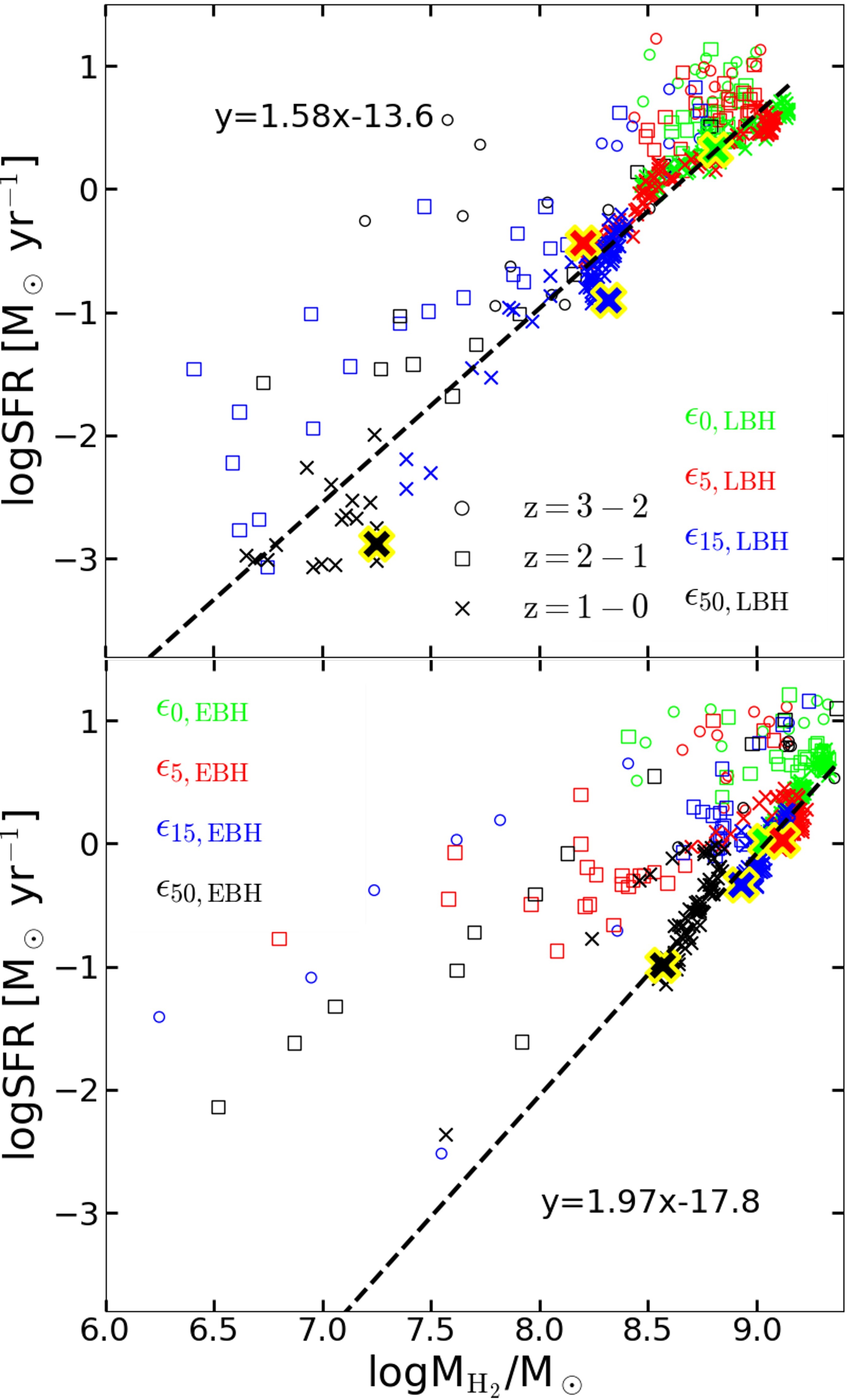}
\caption{SFR versus molecular mass, $M_{\rm H2}$, within the galactic radius $0.1R_{\rm vir}$ for LBH models (top frame) and EBH models (bottom frame). Points represent each 100\,Myr of evolution during $z=3-0$. Those corresponding to $z=0$ for each model are highlighted with a yellow outline. Marker color shows the accretion efficiency: lime green for $\epsilon_{0}$, red for $\epsilon_{5}$, blue for $\epsilon_{15}$, and black for $\epsilon_{50}$. Marker type indicates time, namely, circles correspond  $z=3-2$, squares to $z=2-1$, and x's to $z=1-0$. The dashed lines are median lines which provide linear fits for $z\ltorder 1$ markers. The markers converge to median lines after $z\sim 2$. 
\label{fig:SFRvsfh2}}
\end{figure}

In the lower frame of Figure\,\ref{fig:SFR_comp}, we examine the two $\epsilon_0$ models which only differ in their SN feedback strength.  In this case we observe even fewer changes to the SFR evolution. The differences are small, except at $z\sim 1$ and for $z > 2$, and the final $M_*$ differ by less than 7\%.

As the gas content in galaxies is expected to play a major role in governing the SF, we turn to the evolution of the gas fraction there, which is defined as the gas fraction of the total baryonic mass within $0.1R_{\rm vir}$. Figure\,\ref{fig:fgas} exhibits the evolution of the gas fraction, f$_{\rm gas}$, in all models. We can distinguish roughly two regimes in evolution, namely, $z\gtorder 2$ and $z < 2$. The earlier one is characterized by high f$_{\rm gas}$, up to $\sim 0.6$. During this time, f$_{\rm gas}$ is strongly variable, but declining. The latter regime displays much lower values, f$_{\rm gas}\ltorder 0.2$, and is much less variable, apart from $\epsilon_{\rm 50,LBH}$ which has a nearly empty cavity. Between $z\sim 2-1$, the gas fraction becomes negligible in most of the AGN models, and the SFR also decreases. On the average, f$_{\rm gas}$ in the EBH sequence is higher than in the LBH models after $z\sim 2$.

Because the strongest decreases in the SFR often correspond to major reductions in f$_{\rm gas}$, we have attempted to correlate both quantities. Significant drops in f$_{\rm gas}$, below 5\%, have been analyzed, and found to result from a combined effect of SN and AGN feedback working in tandem with interactions, mostly with the halo substructure. In some models, e.g., $\epsilon_{\rm 5,EBH}$, where f$_{\rm gas}$ drops below a few percent, we observe destruction of the gaseous disk and its re-formation in the polar plane with respect to the stellar disk, especially during $z\sim 2.5-1$. We detect a substantial SFR in this polar disk, which remains for a few Gyr. The action of the AGN in most cases exacerbates the gas loss in galaxies, which is pushed out beyond $0.1R_{\rm vir}$.

However, we do not find a good correlation between the SFR and f$_{\rm gas}$. For lower f$_{\rm gas}$, the SFR varies by $\sim 3$ orders of magnitude, and for high f$_{\rm gas}$, the SFR appears independent of the gas fraction. We, therefore, turn to the molecular gas content, and attempt to correlate it with the SFR. To make this dependence more transparent, we replace the molecular gas fraction with molecular gas mass, $M_{\rm H2}$. The reason for this is straightforward --- cold H$_2$ gas increases clumpiness, and, because the SF depends on the local density, correlation between the SFR and molecular gas is expected.

\begin{figure}[ht!]
\center
\includegraphics[width=0.9\linewidth]{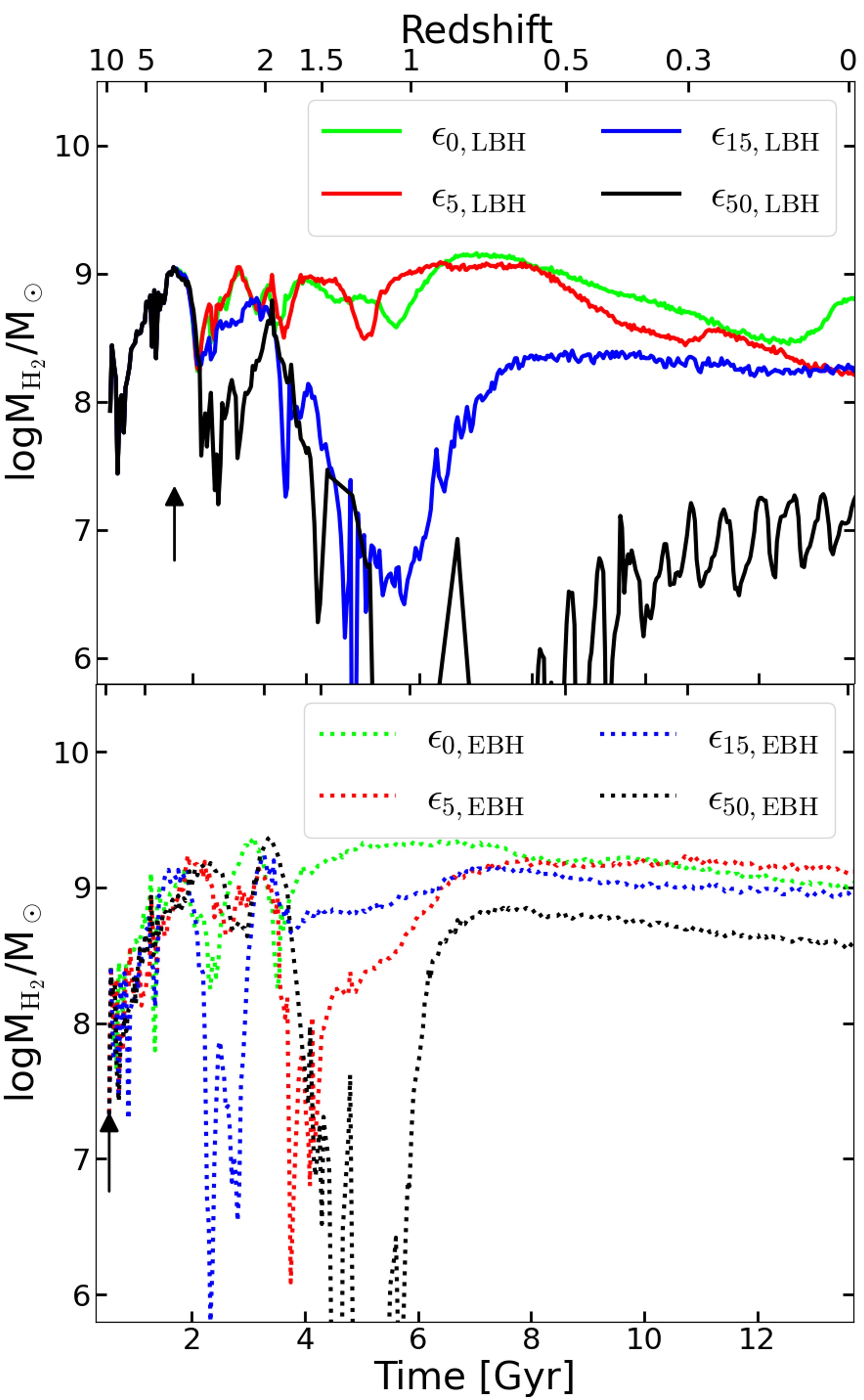}
\caption{Evolution of $M_{\rm H_2}$ inside $0.1R_{\rm vir}$, for LBH (top) and EBH (bottom). The black arrows indicate the seeding time of the SMBHs.
\label{fig:Mh2}}
\end{figure}

Figure\,\ref{fig:SFRvsfh2} displays a dependency of SFR on $M_{\rm H_2}$. The fit-line displayed is a linear least-squares fit to all data points after $z=1$. The slopes appear similar, but are slightly steeper for the EBH models. The LBH models with high AGN feedback have lower $M_{\rm H_2}$ for prolonged time periods, unlike in the EBH model. This is clear in the time evolution of $M_{\rm H_2}$ shown in Figure\,\ref{fig:Mh2}. The resulting SFR--$M_{\rm H_2}$ correlation is not tight but shows up clearly. This relation can be connected to the evolution of the CO distribution and to the IR galaxy luminosity function \citep[e.g.,][]{daddi13}, but is outside the scope of this work. 

\subsection{Evolution of the SMBH and its Feedback}
\label{sec:SMBH}

\begin{figure}[ht!]
\center
\includegraphics[width=0.9\linewidth]{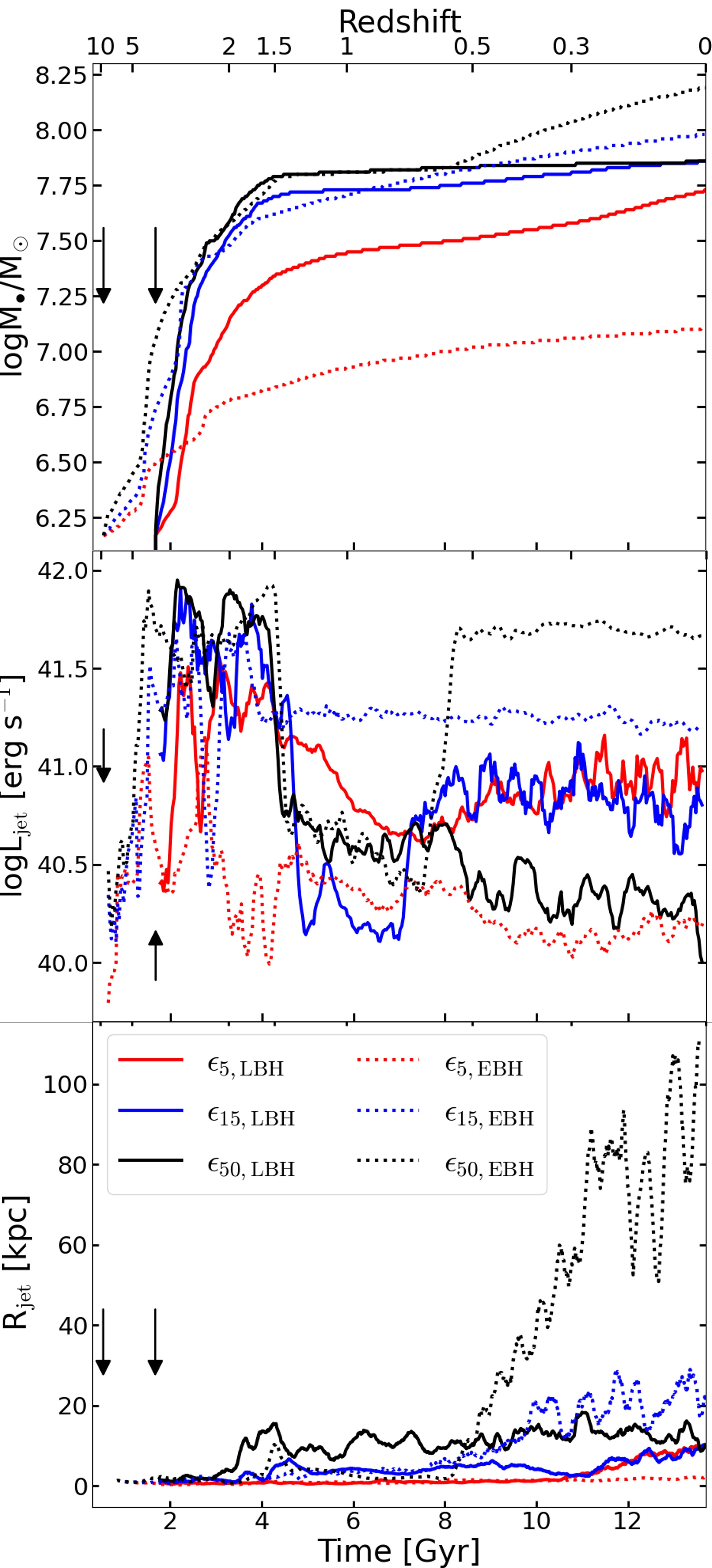}
\caption{Evolution of the SMBH mass, $M_\bullet$ (top), the total luminosity of the jets, $L_{\rm jet}=\eta[1/2 \dot{M}_\bullet v_{\rm jet}^2 + 3/2 (\dot M{_\bullet}/m_p)kT] $ (middle), and evolution of jet extent, R$_{\rm jet}$. R$_{\rm jet}$ is represented by the median distance of the farthest 20\% of the jet particles at a moment in time, sampled every 30\,Myr. The black arrows indicate the seeding time of the SMBHs.
\label{fig:BHmass}}
\end{figure}

Next, we turn to the SMBH evolution in our simulated galaxies. Figure\,\ref{fig:BHmass} shows the growth of $M_\bullet$ in the top frame, and the total luminosity of the jet, L$_{\rm jet}$, in the middle frame. For the LBH seeds, initial extremely rapid growth is replaced by a very slow growth after $z\sim 2-1.5$. The $\epsilon_{\rm 15,LBH}$ curve becomes nearly flat for about 4\,Gyr, then starts to grow again as the galaxy recovers gas near the center. The $\epsilon_{\rm 15,LBH}$ and $\epsilon_{\rm 50,LBH}$ models lie very close at all times and trend to the same value at $z=0$, $M_\bullet\sim 7.2\times 10^7\,M_\odot$. The $\epsilon_{\rm 5,LBH}$ model lies below by a factor of $\sim 2-3$, but ends up only about a factor of $1.5$ below the other LBH models, as the accretion rate for this model is the highest of the LBH AGN models after $z\sim 1.5$. 

This evolution differs for the EBH seeds. The growth rate is slower for $z\gtorder 2$. After this, the slope of the growth rate remains higher than for the LBH models, and after $z\sim 1$ even accelerates, except for $\epsilon_{\rm 5,EBH}$. While $\epsilon_{\rm 15,EBH}$ and $\epsilon_{\rm 50,EBH}$ curves lie close until $z\sim 0.6$, they end up separated by a factor 1.8 at $z=0$, at $M_\bullet\sim 0.8-1.5\times 10^8\,M_\odot$. However, the $\epsilon_{\rm 5,EBH}$ growth curve lies well below the other models at all times, and ends at $M_\bullet\sim 10^7\,M_\odot$ only. Hence, the final masses of the SMBHs differ by about a factor of 10, which requires additional explanation to AGN feedback --- this is explored below and in section\,\ref{sec:discuss}.

The AGN feedback can be the primary cause affecting the SMBH growth, of course in tandem with the gas supply to the inner regions of a galaxy. This feedback can be divided into direct interactions of the jet particles with the ambient gas, and into development of expanding bubbles triggered by the jet energy deposition. The bottom of Figure\,\ref{fig:BHmass} shows $R_{\rm jet}$, defined here as the median distance, relative to the galaxy's potential minimum, of the farthest 20\% of jet particles at a moment in time. This radius can be interpreted as the average range of jet energy deposition in the ISM. The extension of jets, $R_{\rm jet}$, correlates with $\epsilon$. For most models, jets have been contained within $\ltorder 20$\,kpc from the galaxy center. Figure\,\ref{fig:BHmass} reveals that after $z\sim 0.5$, when $L_{\rm jet}$ consistently exceeds $\sim 5\times 10^{40}\,{\rm ergs\,s^{-1}}$, the corresponding $R_{\rm jet}$ evolution tends to have mostly a positive slope. This suggests that the jet is able to `break-out' at later times provided it can reach this critical luminosity. 

The two models for which this does not occur, $\epsilon_{\rm 50,LBH}$ and $\epsilon_{\rm 5,EBH}$, have R$_{\rm jet}$ values which remain relatively flat at late times. However, for the $\epsilon_{\rm 50,LBH}$ model, R$_{\rm jet}$ is about 10 times that in $\epsilon_{\rm 5,EBH}$, despite comparable  $L_{\rm jet}$ values. This difference in $R_{\rm jet}$ is related to the lack of gas in the central region of the $\epsilon_{\rm 50,LBH}$ galaxy, and the jet particles are able to propagate much farther unimpeded, but are scarce due to low $L_{\rm jet}$, which is in turn caused by the lack of gas and associated low $\dot M_\bullet$ in the central region.  

We define the stellar disk rotation $z$-axis based on the inner 3\,kpc. Also, in general, we find that, in the central 3\,kpc of the galaxy, the stars and gas are co-planar, except when the gas is strongly perturbed by interactions. But at larger radii, we often have accretion of gas that is misaligned with the stellar disk midplane, which is eventually funneled towards the inner disk, as well as outflowing gas (see discussion in section\,\ref{sec:morph}). Jet particles can encounter the gas in both of these regimes. We have looked for correlation between the jet polar angle and the SFR within the galaxies. However, the jet particles are quickly de-collimated and scattered away from their injection direction, especially those that are trapped within the inner few kpc. Also, the SFR is affected by the expanding bubbles, whose properties we address in the next section.

\begin{figure}[ht!]
\center
\includegraphics[width=1.0\linewidth]{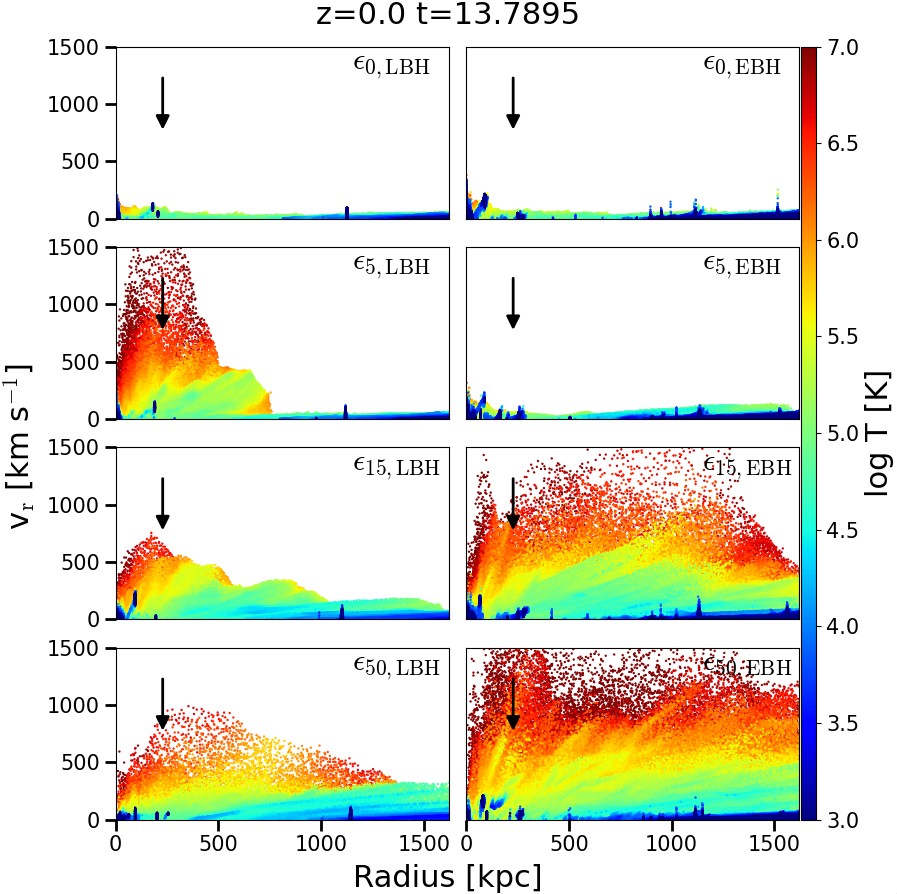}
\caption{Evolution of radial velocity of gas particles (and shocks) versus radius (in comoving coordinates, divided by $h$), describing the expanding bubbles, with color representing temperature. The black arrows show the position of R$_{\rm vir}$ on the $x$-axis, R$_{\rm vir}$ is shown as a function of time in the associated {\it Animation}. Bubbles are indicated by the propagation of hot outflowing gas. The linked animation shows this Figure for $z=9-0$ and starts with a detailed description of the axes, colors, units, etc.
\label{fig:VRad}}
\end{figure}

\subsection{Secondary feedback from the SMBH: expanding bubbles}
\label{sec:cocoons}

While the jet particles do not typically propagate more than a few tens of kiloparsecs, their effects are seen on Mpc scales.  The jet particles interact hydrodynamically with the surrounding gas and deposit energy by heating the gas and generating shocks that propagate into the CGM and beyond, into the IGM.  Inside these expanding `bubbles', the gas is very hot, $T > 10^{6-7}$\,K, and expands with a sound speed up to $\sim 3,000\,{\rm km\,s^{-1}}$. So the bubble expansion is supersonic with respect to the downstream gas which is much colder. 

In order to analyze the propagation of these jet-generated bubbles, we have plotted their positive (outward) radial velocities versus radius for all the gas particles that lie within 1.6\,Mpc of the galaxy center (see Fig.\,\ref{fig:VRad}). Here we color each gas particle by its temperature and can identify bubbles by concentrated regions of hot, fast, outflowing gas.  This Figure links to an animation showing the bubble propagation from $z\sim 9$ to $z=0$, for all our models. Velocities of the shock fronts are represented by the motion of discontinuities advancing along the $x$-axis.  We find that shock velocities range from a maximum value of $\sim 3,000\,{\rm km\,s^{-1}}$ down to $\sim 200\,{\rm km\,s^{-1}}$.  Faster shocks generally correspond to higher $\epsilon$, high $z$, and the initial outbreaks of the bubbles. The bubbles frequently overlap and the later ones re-energize the expansion.  The SN feedback can stir the gas within $\sim R_{\rm vir}$, as can be seen in the $\epsilon_0$ comparison models.  
 
From the animation in Figure\,\ref{fig:VRad}, we have determined that gas particles inside the bubble are at their highest velocity and temperature when they leave the galaxy. They cool and slow down, especially outside $R_{\rm vir}$.  When the gas interior to the shock reaches the shock front which delineates the bubble, it quickly decelerates and heats up as its kinetic energy is converted to thermal.  This can be seen as cascading gas particles become redder as they reach the discontinuities.  The frequency of occurrence  of new bubbles is higher at early times, $z\gtorder 1.5$, when the accretion rates onto the galaxy and onto the SMBH are higher.   

Bubbles form within $\sim 300$\,Myr of planting the SMBH seeds in all AGN models, except in $\epsilon_{\rm 5,LBH}$. Therefore, the EBH models exhibit bubbles $\sim 1$\,Gyr earlier than the LBH models. The $\epsilon_{\rm 5,LBH}$ model is unique among the AGN models as it does not form any significant bubbles until $z\sim 0.25$.  The reason for this delay will also be discussed in section\,\ref{sec:discuss}. Only once the gas density substantially drops around the galaxy, as in the $\epsilon_{\rm 50,EBH}$ model after $z\sim 0.5$, do we see the jet particles remaining collimated to significant distances from the galaxy. Otherwise, the jet particles quickly de-collimate and deposit energy nearly isotropically in and around the galaxy. The bubbles tend to have axial ratios between 1:1 and 3:1, and become more anisotropic when they breakout of the cosmological filaments, where the parent DM halos are embedded. We do observe multiple nested bubbles which are misaligned relative to each other, when the SMBH axis is `rapidly' changing due to the angular momentum of accreted matter. Regardless of the behavior of the jet particles, the bubbles tend to propagate along the path of lowest thermal pressure outside R$_{\rm vir}$. 
  
To summarize, the effects of the bubbles on the galaxy and its environment are multivariate and substantial.  We see changes to galaxy metallicity distribution (see Paper\,I), gas content (and thus the SF), and morphology. In addition, the chemical and thermodynamic properties of the CGM are greatly influenced.  These are presented and discussed further in the remainder of the results section, as well as in the discussion section. 

\subsection{Evolutionary effects in quantitative galactic morphology}
\label{sec:morph}

\begin{figure*}[ht!]
\includegraphics[width=1.\linewidth]{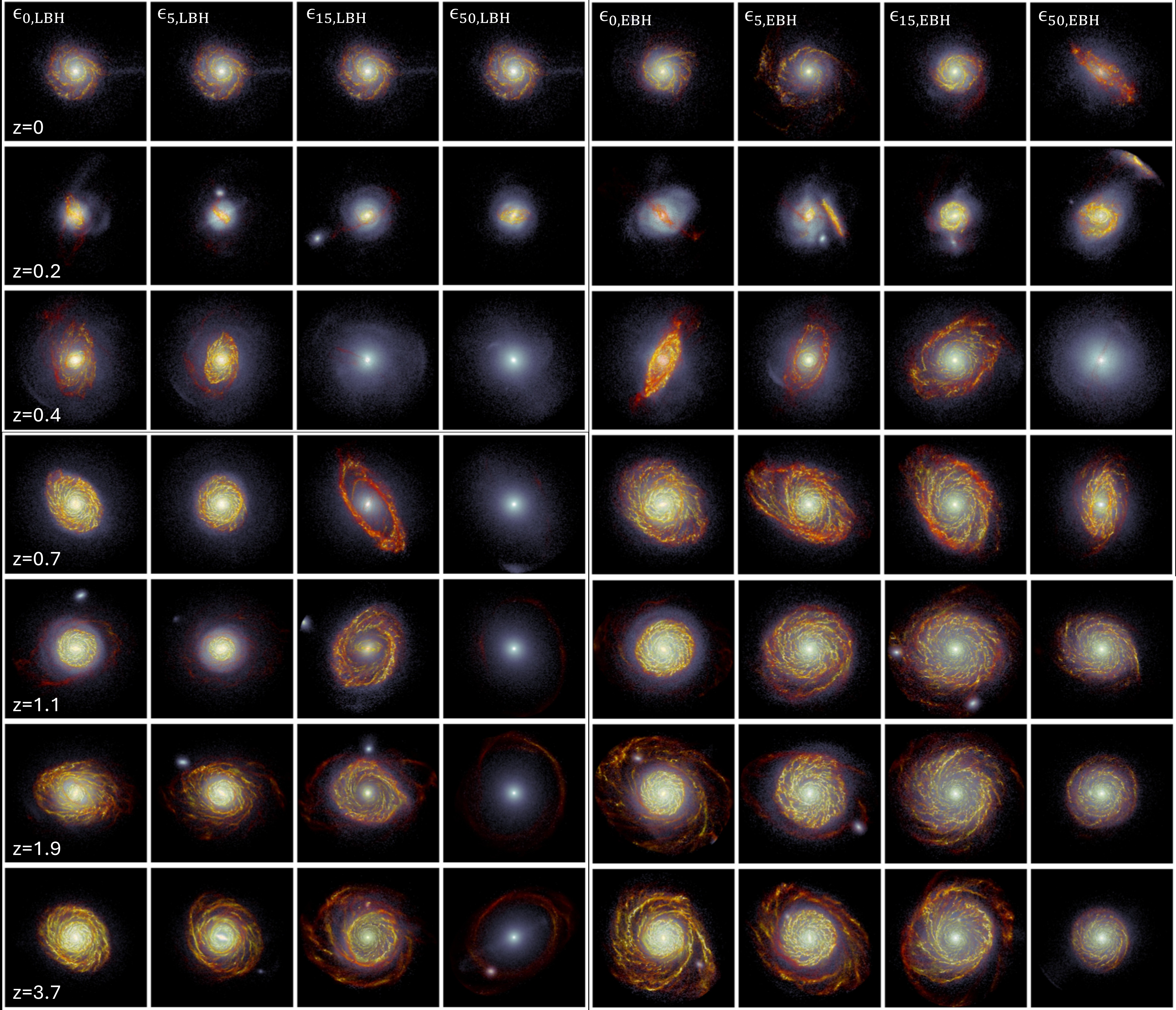}
\caption{Face-on galaxies in 60\,kpc x 60\,kpc frames from LBH and EBH sequences shown for each 2\,Gyr from $z=3.7$ to $z=0$. Stars are blue, and gas is over-plotted on the stars in orange, both colored by 2-D projected density and rotated based on the angular momentum vector of stars inside 3\,kpc. The first 4 columns represent the LBH sequence galaxies, from $\epsilon_0$ to $\epsilon_{50}$, and the next 4 columns represent the EBH sequence.
\label{fig:galaxies}}
\end{figure*}

\begin{figure*}[ht!]
\includegraphics[width=1.\linewidth]{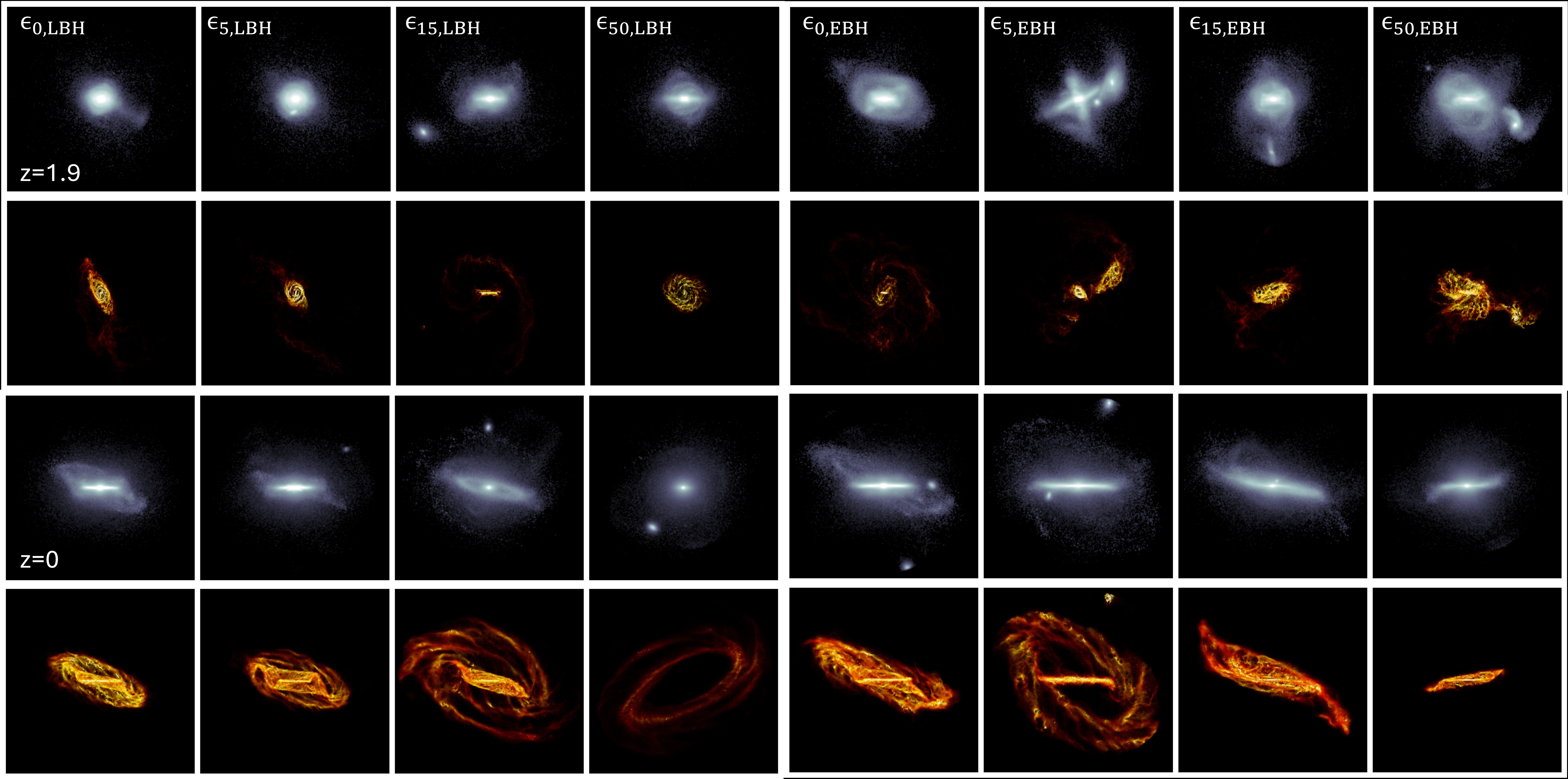}
\caption{Galaxies from Figure\,\ref{fig:galaxies}  shown edge-on with LBH (left 4 columns) and EBH (right 4 columns) sequences shown at $z=1.9$ and $z=0$. Stars  (first and third rows) and gas (second and fourth rows) are separated . Stars are blue, and gas is orange. Both are colored by 2-D projected density and rotated based on the angular momentum of stars inside 3\,kpc. Frames are 60\,kpc x 60\,kpc. 
\label{fig:galaxies2}}
\end{figure*}

The results discussed in the previous sections display a complex and sometimes dramatic evolution of modeled galaxies subject to SN and AGN mechanical feedback. Evidently, the resulting galactic morphologies change with redshift. Some of the galactic properties can be easily quantified, such as stellar and DM masses, gas fraction, SFRs, SMBH masses and accretion rates, etc., others are more challenging to quantify. Figure\,\ref{fig:galaxies} displays the evolving stellar and gas face-on morphologies of simulated EBH and LBH galaxy sequences. This section attempts to characterize the most important changes in stellar and gas morphologies, and this Figure provides a visual overview of this evolution. Note that the DM halo has been assembled by $z\sim 2$.

Figure\,\ref{fig:galaxies} displays the galaxies for $z\sim 3.7-0$, when both sequences can be directly compared. This figure confirms that evolution of the stellar and gas disks with redshift is not monotonic due to the combined effects of interactions and SN around $z\sim 2.5-1$ as explained in section\,\ref{sec:SF}. This is particularly evident in the $\epsilon_{\rm 15,LBH}$ and $\epsilon_{\rm 50,LBH}$ models, these galaxies are visibly more gas poor --- this correlates well with the substantial decrease in their SFRs. On the other hand, models without AGN, display a more monotonic disk growth. For example, the $z=0$ models in both LBH and EBH sequences show more extended disks, from $\epsilon_0$ to $\epsilon_{15}$, and abruptly smaller or absent  gas disk for $\epsilon_{50}$ models. 

The $\epsilon_{\rm 50,LBH}$ model, which loses most of the gas by $z\sim 2$, displays visible evolution in its stellar disk, which becomes more centrally concentrated. Indeed, comparing the gas surface density at $z=0$, we find that the $\epsilon_0$ galaxies have the highest central densities and lowest densities at $0.1R_{\rm vir}$. While $\epsilon_{15}$ and $\epsilon_{50}$ galaxies have the highest `edge' densities (e.g., Fig.\,3 of Paper\,I). This increase in the edge gas surface densities at $z=0$ can be also seen in Figure\,\ref{fig:galaxies2}, especially in the EBH galaxies. Hence, the gas distribution is more extended in our EBH models.  The exception  is the $\epsilon_{\rm 50,EBH}$ model which apparently lost much of its gas. The simple explanation for this effect is related to the AGN feedback which plays a dual role --- pushing out the ISM and acting to prevent the inflow from the CGM. This issue is further explored below and in section\,\ref{sec:discuss}.

We find it instructive to present a sub-sample of galaxies from Figure\,\ref{fig:galaxies} in a different projection, edge-on at $z=1.9$ and $z=0$ (Fig.\,\ref{fig:galaxies2}). For clarity, we have separated the stars from gas. While all galaxies appear as disks, both in stars and in gas, these images underline the fact that the inflowing gas has a distinctly different angular momentum compared to the galactic stellar disks. Not only does the accreted gas form inclined rings, but these gas rings are accompanied by SF. With time the gas settles down to the stellar disk midplane, while the relic stellar rings appear to be more long-lived.

To quantify the galactic morphology, we carried out two methods to decompose the galaxy into bulge and disk components.  Similarly to Paper\,I,  we carried out the Sersic decomposition for each 1\,Gyr of evolution from $z=3.7$ to $z=0$.  This analysis revealed that a double exponential profile represented the best fit to stellar surface density, $\Sigma_{*}$, profiles, not only at $z=0$, but several times throughout the evolution of the galaxies.  Double exponential galaxies are noted in observations and simulations \citep[e.g.,][]{pohlen06,erwin08,struck19}. We found that the formation of outer disks corresponds to times when misaligned gas is accreted onto the galaxy and forms stars in a polar\footnote{Strictly speaking this is not a polar ring which forms on a stable polar orbit, but is on a secularly evolving orbit.} ring. Polar ring galaxies are common in our universe and are thought to form as a result of tidal accretion events, similar to what we observe in our simulations, shown in Figure\,1 of Paper\,I \citep[e.g.,][]{sparke86,binney08}. 

Over time, the inner and outer disks become coplanar and merge in our simulations. Gaseous components end up in the stellar disk plane due to dissipation, while the stellar component can be smeared, contributing to a thick disk. Importantly, the misaligned accretion can be in the path of the AGN jet or its expanding bubble, even when the jet is aligned perpendicularly to the galaxy disk.  When this occurs, the AGN feedback disrupts accretion of gas onto the galaxy with significant impacts to galaxy morphology and growth. We see this most obviously during the last $\sim 1$\,Gyr of evolution in the $\epsilon_{\rm 50,EBH}$ model. Here we have jets propagating out to $\sim 50-60$\,kpc, and interact with the gaseous ring and accretion flow. This in turn can minimize the gas accretion onto the galaxy, resulting in the reduced galaxy size and stellar mass, that we see in Figure\,\ref{fig:galaxies}. 

In order to supplement and verify the Sersic decomposition, we have carried out kinematic decomposition for every 30\,Myr of galaxy evolution, following \citet{bi22a}. Using stars within $0.1R_{\rm vir}$ and defining the bulge as the population of stars with $|j_{\rm z}/j_{\rm c}| < 0.5$, where $j_{\rm z}$ and $j_{\rm c}$ are the specific angular momenta around the $z$-axis and the circular one, respectfully. Disk stars have been taken as those with $|j_{\rm z}/j_{\rm c}|\ge 0.5$.  

We have calculated the bulge-to-disk (B/D) ratio found through this analysis, normalized by the same ratio in $\epsilon_0$ model.  After $z\sim 2$, the B/D ratio increases basically in all AGN  models, except $\epsilon_{\rm 5,LBH}$. This increase correlates with the AGN feedback, with $\epsilon_{\rm 15,LBH}$ and $\epsilon_{\rm 50,LBH}$ models seeing the most significant enhancement of B/D, by more than 50\%. Enhancement is smallest for $\epsilon_{\rm 5,LBH}$ model, which ends up similar to $\epsilon_0$ models at $z=0$. During $z\sim 2.5-1$, which includes a series of minor and intermediate mergers, a substantial decrease in B/D can be observed, except in $\epsilon_{\rm 5,LBH}$. 

\begin{figure}[ht!]
\center
\includegraphics[width=0.9\linewidth]{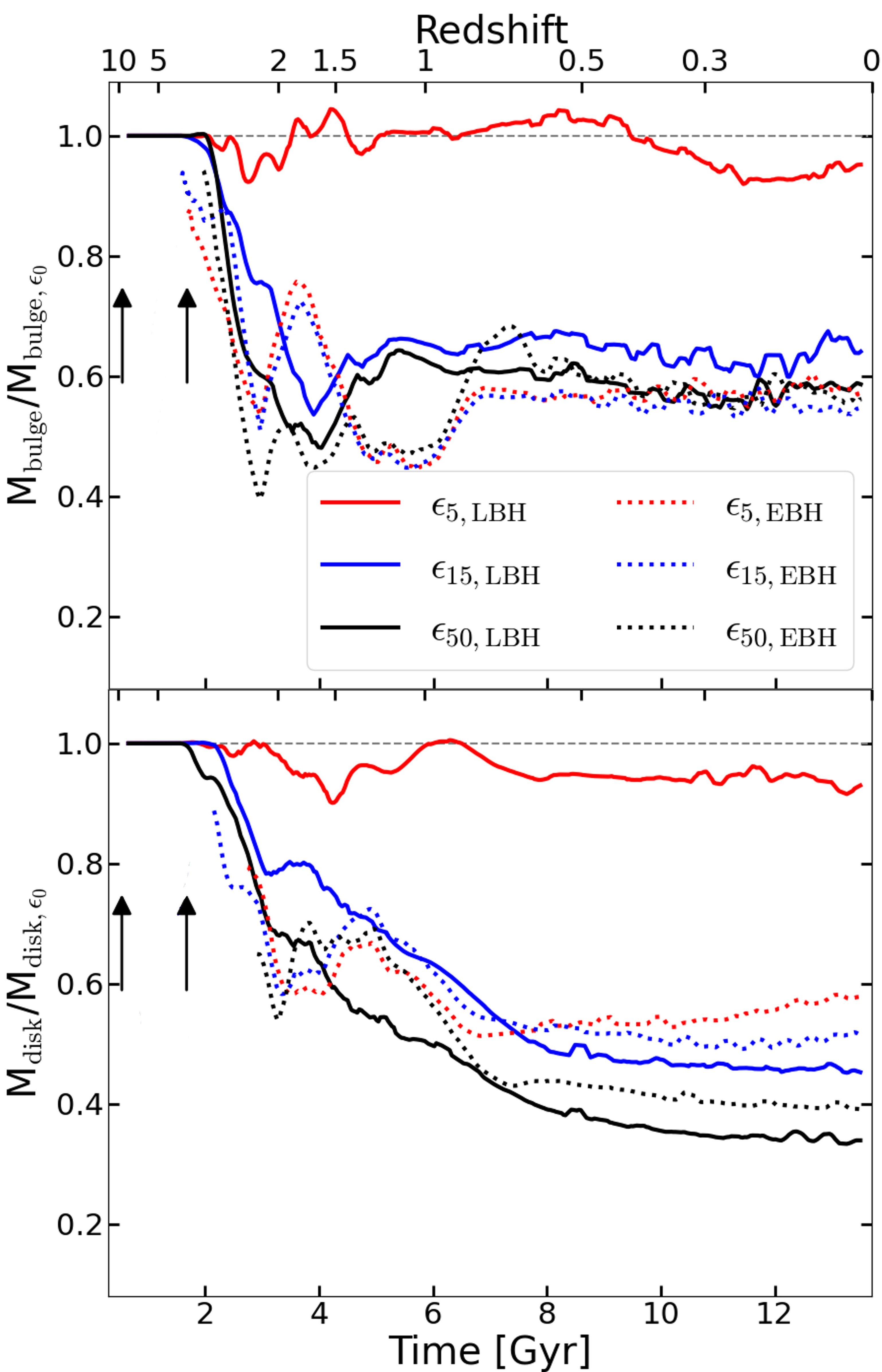}
\caption{Evolution of bulge masses, $M_{\rm bulge}$ (top), and disk masses, $M_{\rm disk}$ (bottom), for the AGN models, normalized by $M_{\rm bulge,\epsilon_0}$ and $M_{\rm disk.\epsilon_0}$.  The straight dashed lines correspond to where this ratio equals unity. The curves for the EBH models are shown only for $z\ltorder 4$, where the kinematic decomposition is reliable. The black arrows indicate the seeding time of the SMBHs.
\label{fig:MassKD}}
\end{figure}

\begin{figure}[ht!]
\center
\includegraphics[width=1.0\linewidth]{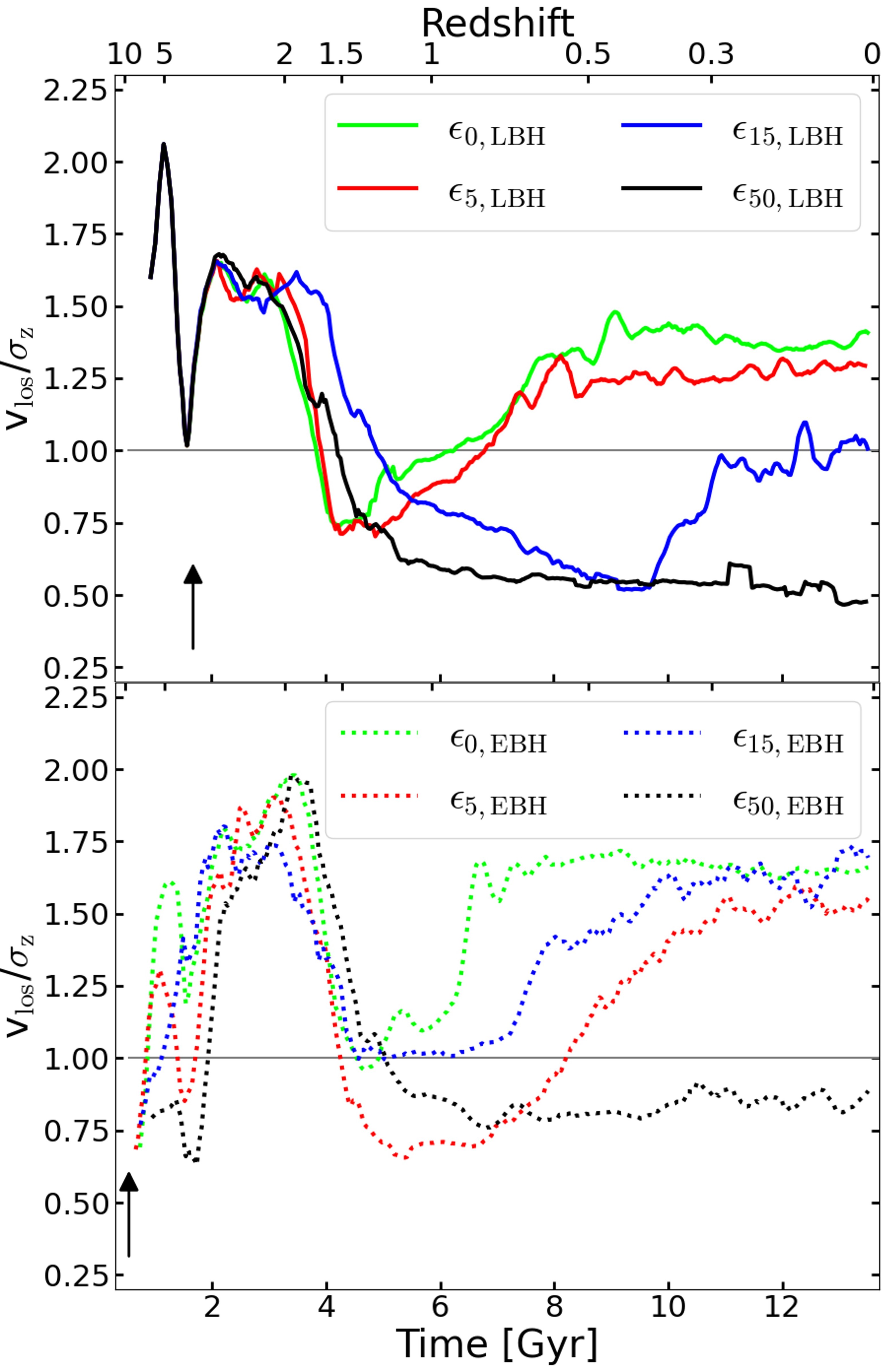}
\caption{Evolution of line-of-sight-to-dispersion velocities ratio, $v_{\rm los}/\sigma_{\rm z}$, in the the LBH (top) and EBH (bottom) sequences. Here, $v_{\rm los}$ has been measured using the maximal value for edge-on galaxies, and $\sigma_{\rm z}$ is the mean z-component of the velocity dispersion, as in \citet{bi22a}.  The arrows indicate the seeding time of the SMBHs.
\label{fig:vlos}}
\end{figure}

To decipher the evolution of B/D, we have separated the evolution of bulges and disks in Figure\,\ref{fig:MassKD}. In fact, both the bulge and disk masses are reduced in the these models, relative to the $\epsilon_0$ models. The bulge reduction is quick after the seeding time coinciding with the development of bubbles in all models, except $\epsilon_{\rm 5,LBH}$ where bubbles develop only at $z\ltorder 0.2$. The bulge masses have been reduced by $\sim 40\%$ and remain flat for the rest of the run time. The associated disk masses have been reduced more gradually but also more substantially, by $\sim 50-70\%$, leading to an overall increase in B/D. 

Hence, we found a direct effect of the developing bubbles on the galaxy SFR as well as the stellar and gas morphologies in AGN galaxies. $\epsilon_{\rm 5,LBH}$, the only model which develops bubbles very late, only after $z\sim 0.2$, evolves similarly to $\epsilon_{\rm 0,LBH}$, as follows from Figure\,\ref{fig:MassKD}. In models where bubbles form, we observe reduction of the gas accretion from the cosmological filaments, and expulsion of a large fraction of gas from galaxies, thus lessening the SFR throughout the galaxy (see sections\,\ref{sec:cocoons} and \ref{sec:discuss}).    

We have compared the results for B/D from Sersic decomposition measured in Paper\,I for the LBH models at $z=0$ with those from the kinematic decomposition obtained here. The stars previously identified as the `inner disk' stars have been found now to be kinematically part of the bulge component. This has affected B/D for two models, namely, $\epsilon_{15}$ and $\epsilon_{50}$, by increasing their B/D ratio. Evidently, the Sersic decomposition based on the stellar surface density should be used with caution.  As indicated previously, we do see formation of outer rings in our galaxies, which ultimately align with the equatorial plane. However, it is not clear from the Sersic profile alone when the transitions occur, leading to altered kinematic states in the various regions of a galaxy.  

Another characteristic measure of rotational support in galaxies is the ratio of the line-of-sight velocity\footnote{{The line-of -sight velocity (v$_{\rm los}$) is the radial velocity with respect to the observer, defined for edge-on stellar disks.}} to the $z$-component of the velocity dispersion, $\sigma_{\rm z}$. Figure\,\ref{fig:vlos} shows the time evolution of the edge-on galaxy line-of-sight velocity, $v_{\rm los}$, in the disk normalized by the $z$-component of the velocity dispersion, $\sigma_{\rm z}$ --- hence, we measure the relative contributions of disk and spheroid. Time periods when $v_{\rm los}$ dominates the galaxy point to a primarily rotational support and indicating the presence of a significant disk component.  We see that, for $z\gtorder 2-1.5$, all models are dominated by the disks, in agreement with \citet{romano-diaz11}. This is followed by a short period when roughly $v_{\rm los}\sim \sigma_{\rm z}$. Thereafter, $\epsilon_{50}$ remains in this state, while other LBH and EBH models return to being disk-dominated. The $\epsilon_{\rm 15,LBH}$ model returns to  $v_{\rm los}/\sigma_{\rm z}\sim 1$ after $z\sim 0.25$. 

Therefore, at low-redshifts our models appear to be more disk-dominated, similar to the $\epsilon_0$ model, except $\epsilon_{50}$ models, which are more spheroid-dominated than $\epsilon_0$, and $\epsilon_{\rm 15,LBH}$ which lies between these regimes. Thus, the difference in evolution of these galaxies can be attributed to the relative strength of the mechanical and thermal feedback from the AGN jets. 

\begin{figure}[ht!]
\center
\includegraphics[width=0.9\linewidth]{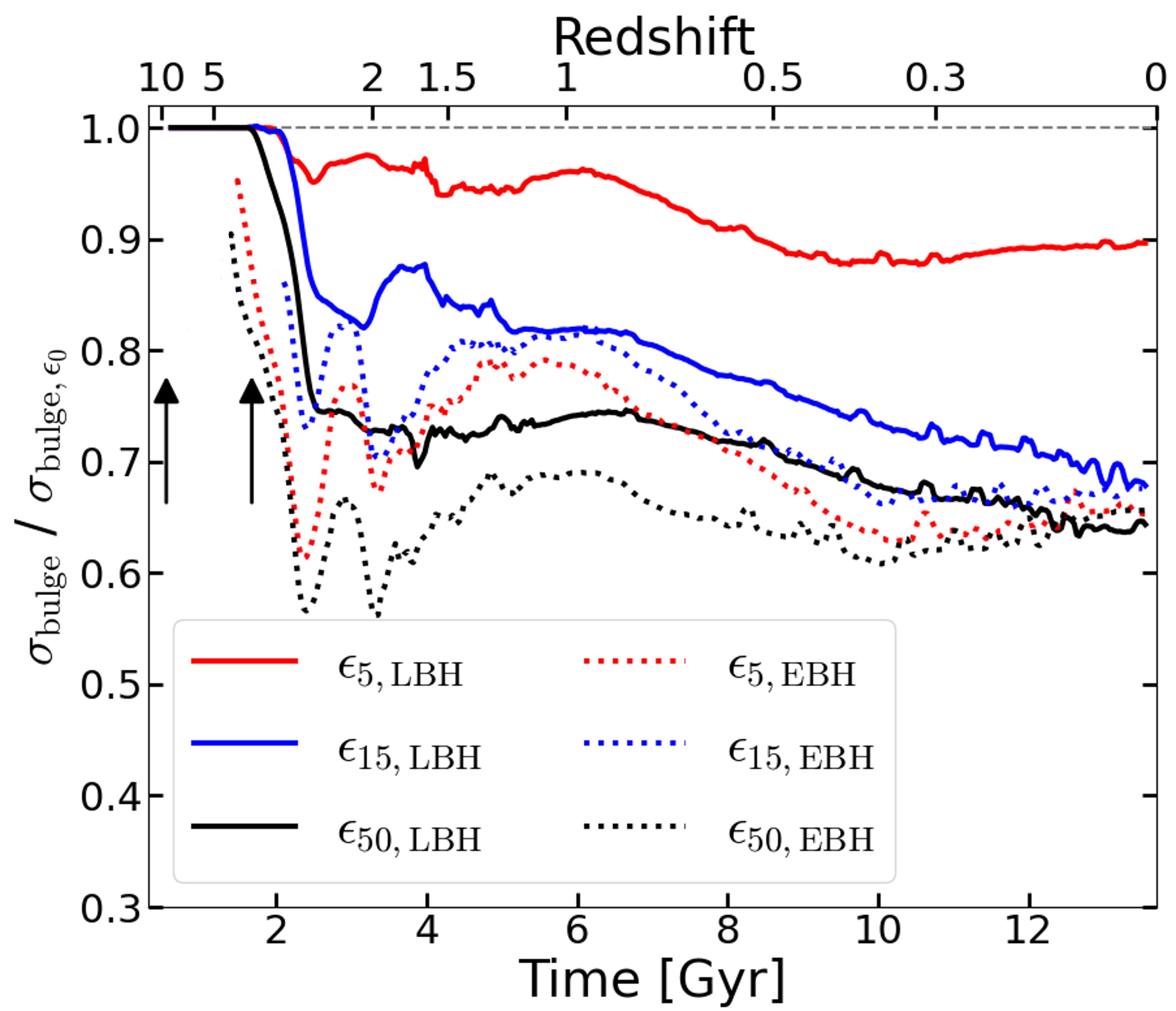}
\caption{Evolution of $\sigma_{\rm bulge}$ in the AGN models, normalized by $\sigma_{\rm bulge,\epsilon_0}$. The dashed horizontal line is for the reference only. As in Figure\,\ref{fig:MassKD} the early evolution of the EBH models has been removed due to noise from interactions. The black arrows indicate the seeding time of the SMBHs.
\label{fig:KD_sigma}}
\end{figure}

To complete discussion about the relative bulge kinematics, Figure\,\ref{fig:KD_sigma} shows evolution of velocity dispersion in the stellar bulge, $\sigma_{\rm bulge}$.  The presence of an AGN decreases $\sigma_{\rm bulge}$ at all times with respect to $\epsilon_0$ model, and the magnitude of this decrease is strongest in models with the highest $\epsilon$.  This is interesting, as one might expect that the presence of an AGN acts to increase $\sigma_{\rm bulge}$ with increasing $\epsilon$, based on the $M_\bullet-\sigma_{\rm bulge}$ relation. The presence of an SMBH in the galaxy center reduces $M_{\rm bulge}/M_{\rm bulge,\epsilon_0}$, as evident from Figure\,\ref{fig:MassKD}, this in turn can reduce $\sigma_{bulge}$. Likewise, Figure\,\ref{fig:SurfDen} shows a reduction in the overall stellar concentration in the central $\sim 1$\,kpc, so `puffing up the bulge.   

\begin{figure}[ht!]
\center
\includegraphics[width=0.82\linewidth]{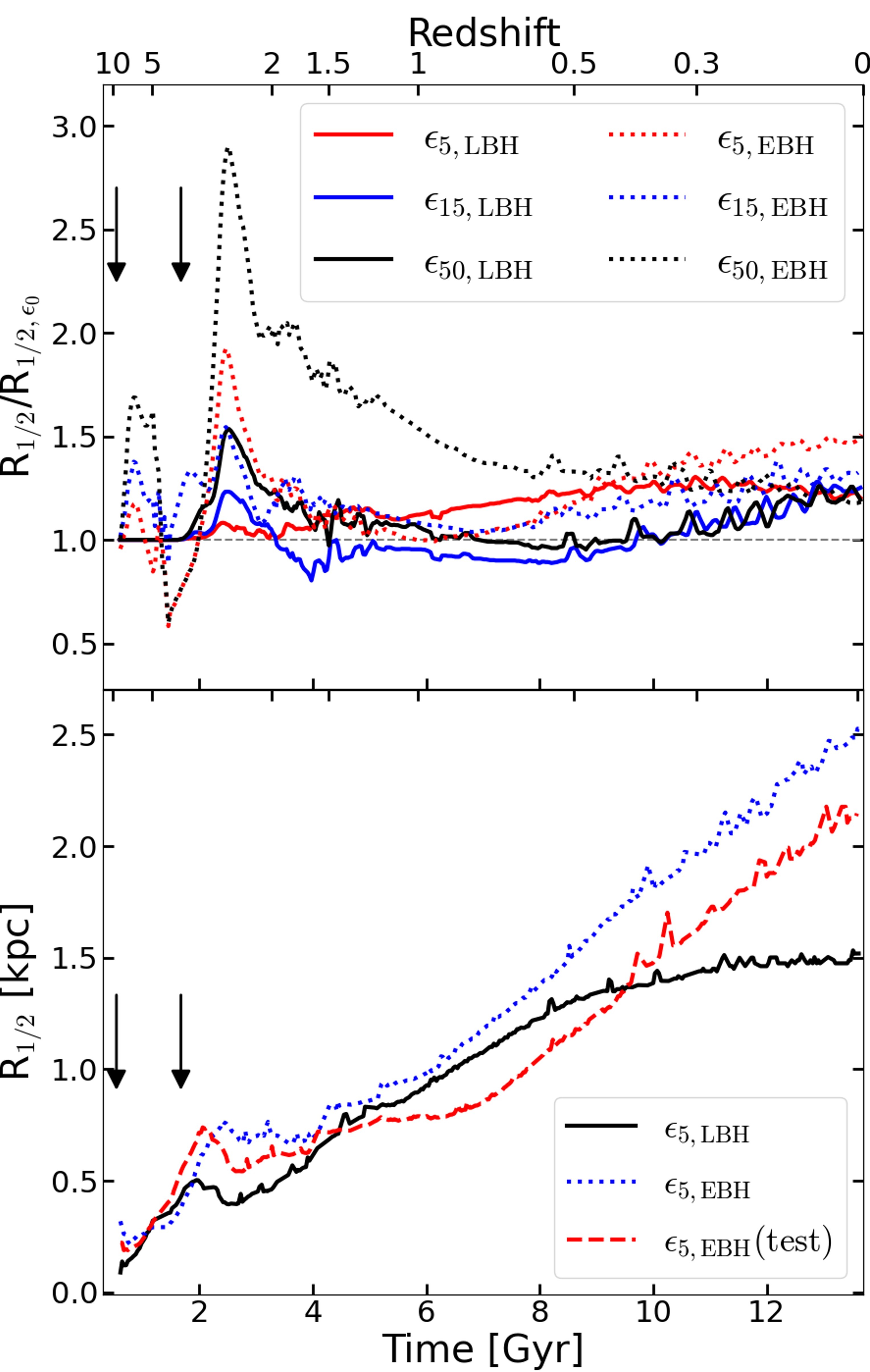}
\caption{Top: evolution of the galaxy half-mass radius $R_{1/2}$ with AGN models normalized by $\epsilon_0$, where $R_{\rm 1/2}$ is defined as the radius enclosing half of the stellar mass residing inside $0.1R_{\rm vir}$. Bottom: three $\epsilon_5$ models showing the effects of seeding early and increasing SN feedback individually. The black arrows indicate the seeding time of the SMBHs.
\label{fig:galR}}
\end{figure}

\begin{figure}[ht!]
\center
\includegraphics[width=1.\linewidth]{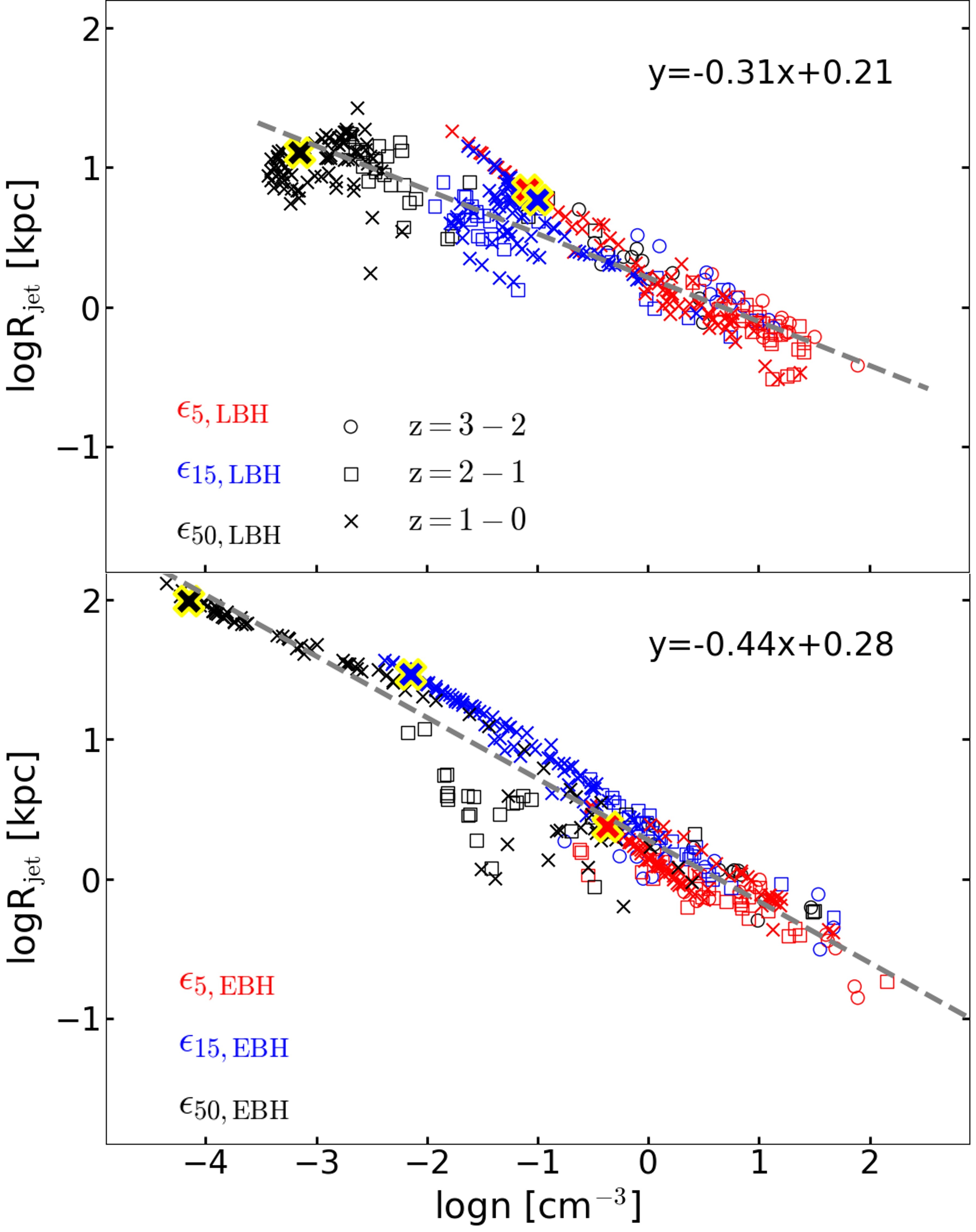}
\caption{$R_{\rm jet}\, - \overline{n}$ correlation. The jet extent $R_{\rm jet}$ versus average gas number density $\overline{n}$ at time $t$ within the spherical volume with a radius $R_{\rm jet}(t)$. The points corresponding to $z=0$ for each model are highlighted with large symbols outlined in yellow. The marker color shows the accretion efficiency: red for $\epsilon_{5}$, blue for $\epsilon_{15}$, and black for $\epsilon_{50}$. The marker type shows the time range, namely, circles correspond to $z=3-2$, squares to $z=2-1$, and x's to $z=1-0$. A linear least-squares fit is over-plotted on the point distribution.
\label{fig:Rjet_ncgm}}
\end{figure}

Consequently, we have analyzed evolution of the stellar mass concentration in the modeled galaxies. Figure\,\ref{fig:galR} (top frame) displays the ratio of the stellar half-mass radius, $R_{1/2}$, in the AGN models normalized by the associated $\epsilon_0$ models.  We observe that the EBH models have $R_{1/2}$ enhanced relative to the $\epsilon_0$ model over most of the evolution time. While the enhanced ratio for $z > 2$ can be attributed to the major merger at $z\sim 4.5$ and related ambiguities in determining this ratio, as well as subsequent minor and intermediate mergers, the prolonged tails for $\epsilon_{\rm 50,EBH}$ and $\epsilon_{\rm 50,LBH}$ models cannot be explained the same way, and must be attributed to the internal evolution in the modeled galaxies. They correspond to a rapid reduction in the SFR for these models, as seen in Figure\,\ref{fig:SFR}. In any case, the relative ratio $R_{1/2}/R_{1/2,\epsilon_0}$ of AGN galaxies ends up elevated by a factor of $1.2-1.5$ at $z=0$. This upward trend in the ratio starts as early as $z\sim 2$ for $\epsilon_{\rm 5,LBH}$ and $z\sim 0.5$ for other galaxies. 

\begin{figure}[ht!]
\center
\includegraphics[width=0.9\linewidth]{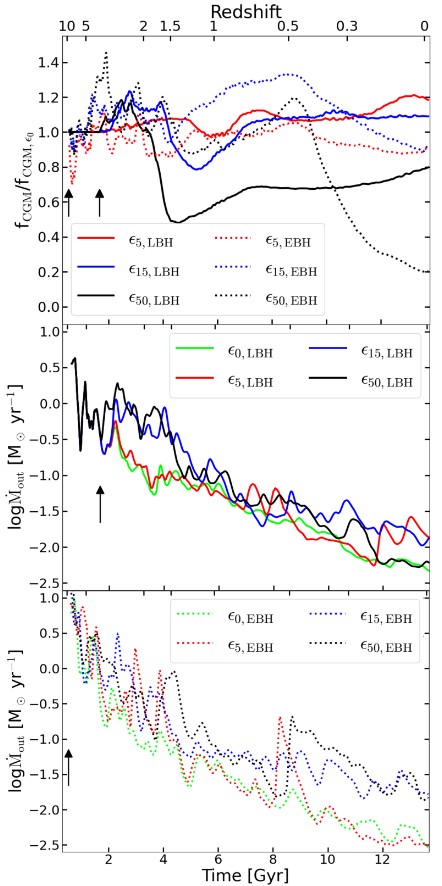}
\caption{Evolution of the gas fraction in the CGM, f$_{\rm CGM}$, in all models (top frame), and outflow across $R_{\rm vir}$, $\dot M_{\rm out}$ for LBH models (middle frame) and EBH models (lower frame). The gas fraction has been determined between $0.1R_{\rm vir}$ and $2R_{\rm vir}$. The black arrows indicate the seeding time of the SMBHs.  
\label{fig:outflow}}
\end{figure}
 
The earlier seeding of the SMBH in the EBH models decreases the mass concentration in galaxies at basically all redshifts, as seen in the lower frame of Figure\,\ref{fig:galR}, comparing the evolution of $\epsilon_{\rm 5,LBH}$ and $\epsilon_{\rm 5,EBH}$. On the other hand, comparing two early seeded models, $\epsilon_{\rm 5,EBH}$ and $\epsilon_{\rm 5,EBH}$(test), where the only difference is in the SN feedback strength, the increased feedback has some marginal effect of $R_{1/2}$.  Finally, comparing $\epsilon_{\rm 5,LBH}$ and $\epsilon_{\rm 5,EBH}$(test) models (with identical SN feedback), we observe multiple crossing of their evolution curves, and the final divergence after $\sim 0.5$, with the early seeded SMBH model ending up less concentrated. 

\subsection{CGM: effects of AGN and SN feedback}
\label{sec:cgm}

The circumgalactic medium (CGM) plays an important role in galaxy evolution being an intermediate between the cosmological flow and the galaxy. This is the region extending  from $R_{\rm vir}\sim 4R_{\rm vir}$ down to the galaxy, where accretion from cosmologcal filaments and diffuse matter is mixed with the galactic outflow from the SN and AGN feedback \citep[e.g.,][]{tumlinson17,bi24}. In this work, we take the outer radius of the CGM as $2R_{\rm vir}$. We have investigated the effect of the jet feedback on the CGM through analysis of its thermodynamic and chemical evolution.   Despite that jets modeled here are mostly trapped within the galaxies and only occasionally does their median extension reach $50-60$\,kpc, we find that jets can have a profound and lasting effect on the properties of the CGM around Seyfert galaxies.

The radial extent of energy deposition by the modeled jets in the CGM is shown in Figure\,\ref{fig:Rjet_ncgm} for EBH and LBH models over the evolution time. Note that $R_{\rm jet}$ increases with decreasing density in the ambient gas. Section\,\ref{sec:cocoons} has shown that the energy deposition by jets has triggered powerful outflows in the form of expanding bubbles, reaching well beyond the galactic ISM and CGM, and even extending deep into the IGM. On scales beyond $\sim R_{\rm vir}$, the expanding bubbles appear to be contained by the cosmological filaments, and breakout perpendicularly to these filaments. 
 
Bubbles are delineated by shocks and push the CGM gas reducing the gas fraction there. As the bubbles clear out gas in the CGM, the jet particles can propagate further before they encounter ambient gas and slow down.  This effect appears to scale with $\epsilon$ such that more efficient accretion, and thus stronger feedback, leads to both a slightly lower density CGM and a jet that can propagate further at later times.

The top frame of Figure\,\ref{fig:outflow} shows evolution of the gas fraction, $f_{\rm CGM}$, within the defined CGM. The EBH and LBH models behave differently --- while the LBH models display a sharp decrease in f$_{\rm CGM}$ after $z\sim 2$ (but not the $\epsilon_{\rm 5,LBH}$ model), the EBH models stay flatter and even increase the gas fraction until $z\sim 0.5$, when a sharp decrease follows.  

Furthermore, we have measured the generated outflow rate, $\dot M_{\rm out}$, calculated from the radially outflowing gas across a 20\,kpc-thick spherical shell at $R_{\rm vir}$ (Fig.\,\ref{fig:outflow}, middle and lower frame). We find that the outflow rate correlates with $L_{\rm jet}$ and f$_{\rm CGM}$, with a delay in the gas response. For example, the sharp decrease in $f_{\rm CGM}$ around $z\sim 2-1.5$ in $\epsilon_{\rm 50,LBH}$ has been preceded by the peak $L_{\rm jet}$ and correlates with the increase in $\dot M_{\rm out}$ at the same time and is followed by a sharp decrease in the galactic SFR in Figure\,\ref{fig:SFR}. The sharp decrease of $\dot M_{\rm out}$ in $\epsilon_{\rm 50,EBH}$ at $z\sim 0.5$ correlates with the sharp increase in $L_{\rm jet}$ occuring at the same time. Despite significant influences to CGM gas content and dynamics, we find that the stellar mass and SFR in the CGM are minimally influenced by the presence of the AGN jet feedback, and small transient effects to these quantities converge back to the no-SMBH values in less than 1\,Gyr.

\begin{figure}[ht!]
\center
\includegraphics[width=1.0\linewidth]{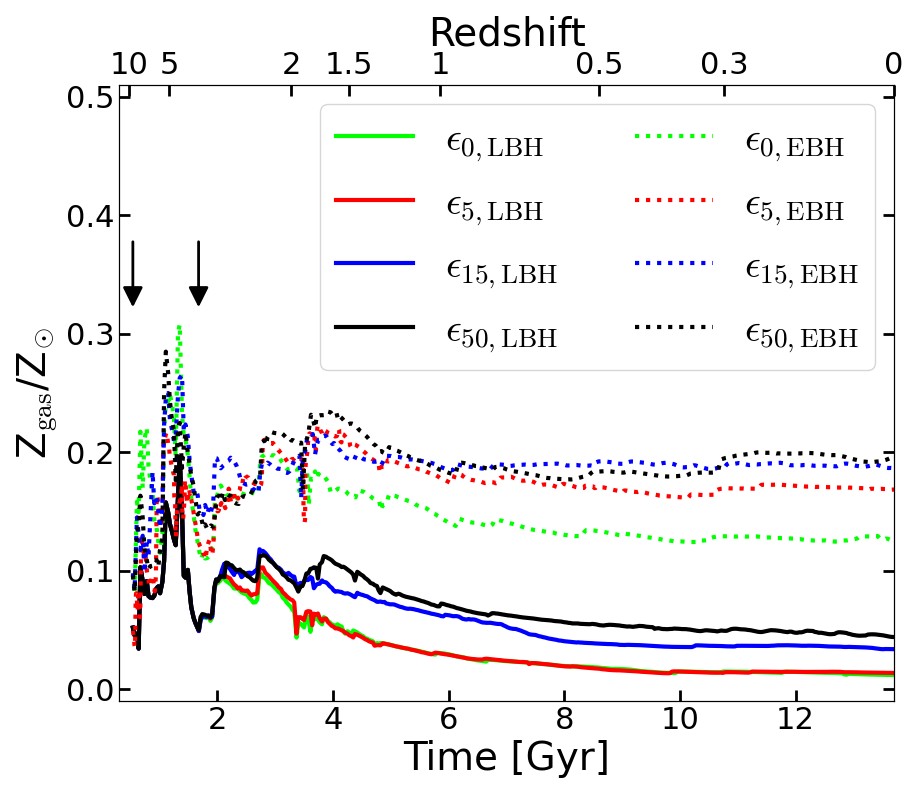}
\caption{Evolution of the gas metallicty, $Z_{\rm gas}$, in the CGM, between $0.1R_{\rm vir}$ and $2R_{\rm vir}$ for all models. The metallicity has been averaged over the CGM volume. The black arrows indicate the seeding time of the SMBHs.
\label{fig:Zcgm}}
\end{figure}

Beyond reduction of the gas fraction in the CGM by increasing the outflow across the virial radius, metallicity of the CGM is affected as well. Figure\,\ref{fig:Zcgm} shows evolution of the gas metallicity in the CGM.  In both the LBH (top) and the EBH (bottom) sequences we see that the AGN models exhibit enhanced metallicity relative to  $\epsilon_0$, basically at all times, except for the $\epsilon_{\rm 5,LBH}$ model. However, this latter model develops bubbles only after $z\sim 0.25$, which explains the lack of enrichment. As analyzed in section\,\ref{sec:SMBH}, the jet-generated bubbles expand into the CGM and carry enriched gas from the central galaxy. They can also strip high metallicity gas from substructures in this region.   

Comparing Figures\,\ref{fig:VRad} and \ref{fig:Zcgm}, we see that sharp increases in $Z_{\rm gas,CGM}$ correspond to strong outflows. It is also evident from Figure\,\ref{fig:Zcgm} that all of the EBH models have $Z_{\rm gas,CGM}$ higher than their associated LBH models starting from $z\sim 9$ --- this early time period is characterized with high $L_{\rm jet}$ and associated expanding bubbles.

\begin{figure}[ht!]
\center
\includegraphics[width=0.9\linewidth]{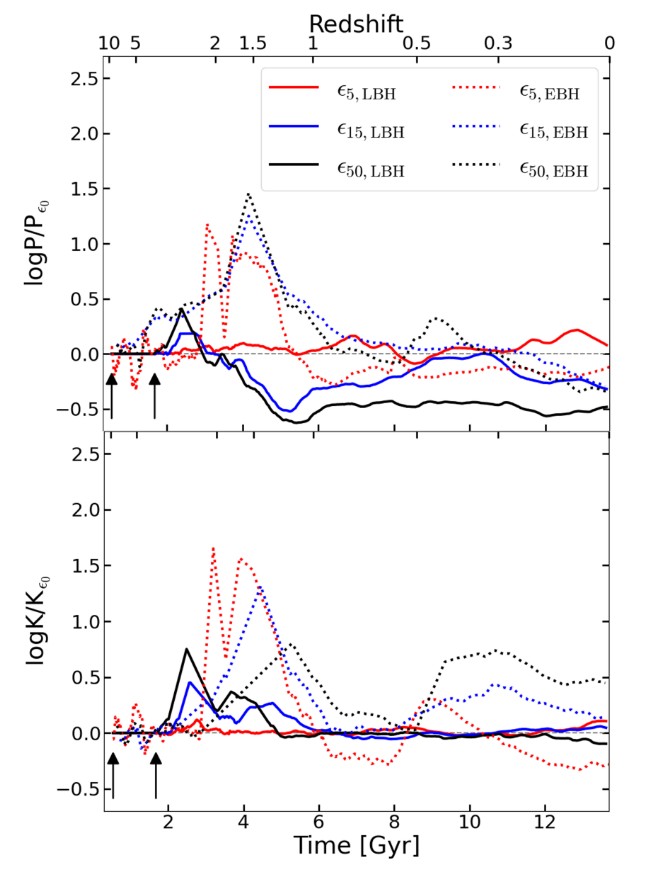}
\caption{Evolution of CGM median pressure and entropy for all gas residing between $0.1R_{\rm vir}$ and $2R_{\rm vir}$. All AGN models are normalized by $\epsilon_0$ in order to show the effect of the AGN feedback. The black arrows indicate the seeding time of the SMBHs.
\label{fig:cgm_props}}
\end{figure}

In order to investigate the source of this relative enrichment in more detail, we have checked the three comparison $\epsilon_5$ models for metallicity evolution, shown in Figure\,\ref{fig:SFR_comp} and discussed in section\,\ref{sec:SF}. For $z\ltorder 2$, we detect a factor of $\sim 8-9$ enhancement in $Z_{\rm gas,CGM}$ for the $\epsilon_{\rm 5,EBH}$ model compared to $\epsilon_{\rm 5,LBH}$, and by a factor of 2 compared to $\epsilon_{\rm 5,EBH}$(test). The EBH model, $\epsilon_{\rm 5,EBH}$, where the SMBH has been seeded earlier and the SN feedback prescription has been boosted relative to $\epsilon_{\rm 5,LBH}$, the early spikes in metallicity at $z\gtorder 4$ have been enhanced compared to the other models. This is due to the formation of larger bubbles in response to both SN and SMBH jet feedback working in tandem, triggering increased outflows as evident in the associated animation with Figure\,\ref{fig:VRad}. The $\epsilon_{\rm 5,EBH}$ model displays the largest enhancement in $Z_{\rm gas,CGM}$ at all times, but we also obtain a factor of $\sim 2$ enhancement in models which only differ by their SMBH seeding time.  

We also compared the $\epsilon_0$ EBH and LBH models, where the only change is due to the SN feedback being stronger in the EBH model. In this case, we see an increase in $Z_{\rm gas,CGM}$ by a factor of $\sim 5$, that is established already by $z\sim 2$, in the model with enhanced SN feedback. We conclude, therefore, that while the bubbles are primarily responsible for spreading metals to large distances, and thus their number and power are directly related to the CGM enrichment, the strength of the SN feedback makes non-negligible contributions to this process as well. This becomes clear when comparing Figures\,\ref{fig:Zcgm} and \ref{fig:VRad}.  The enhanced SN feedback affects mostly the gas within $R_{\rm vir}$ before $z\sim 2$, leading to $Z_{\rm gas,CGM}$ enrichment.   

\begin{figure*}[ht!]
\center
\includegraphics[width=0.9\linewidth]{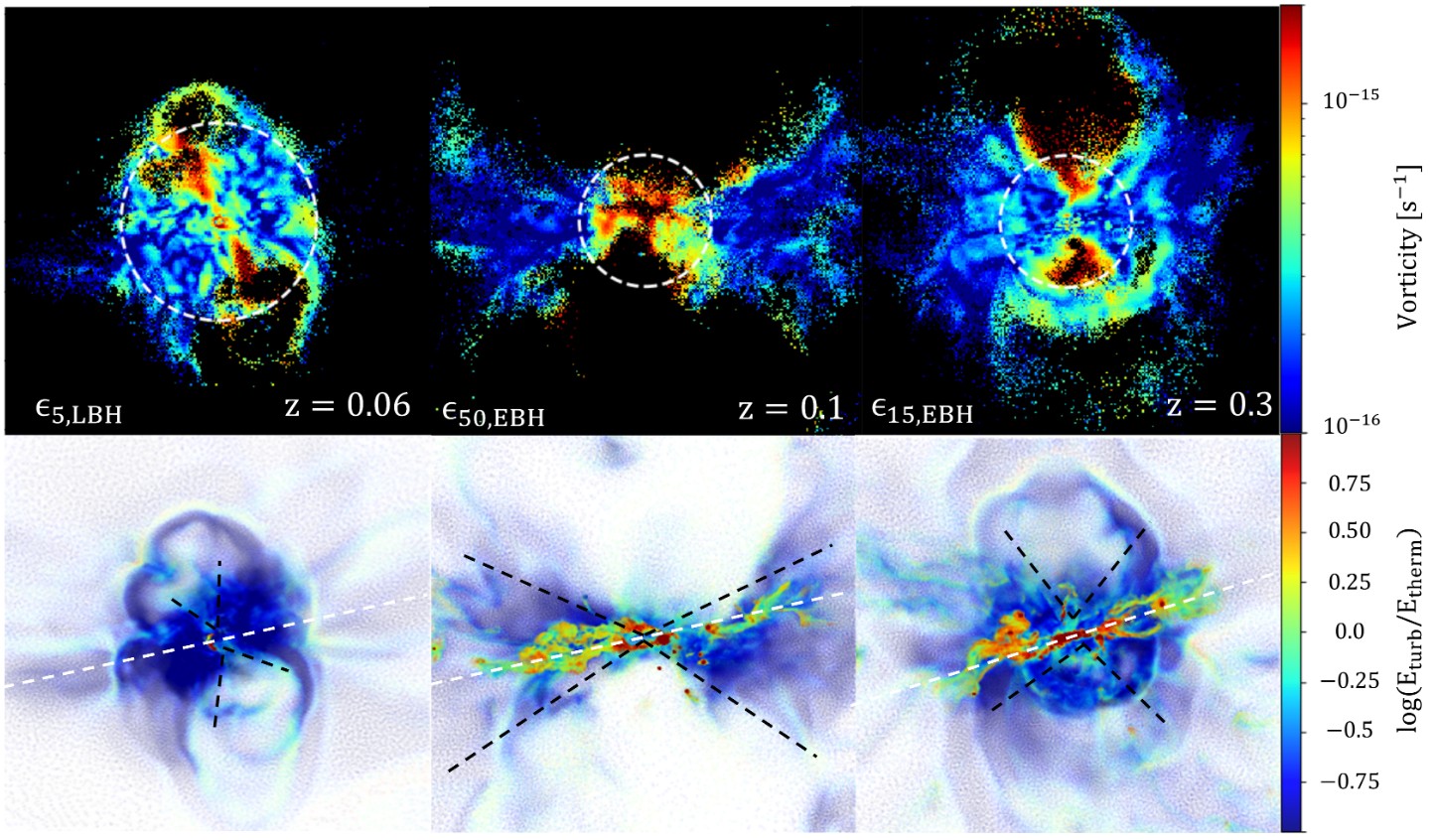}
\caption{{\it Top row:} three examples of CGM vorticity amplitude maps, $|\vec\omega| = |\vec\nabla\times \vec v|$, at $z = 0.06$, 0.1 and 0.3 redshifts displaying bubbles.  Left column shows $\epsilon_{\rm 5,LBH}$ in 1\,Mpc $\times $ 1\,Mpc $\times $ 0.04\,Mpc slices. Center shows $\epsilon_{\rm 50,EBH}$, and right is $\epsilon_{\rm 15,EBH}$ --- both in 1.5\,Mpc $\times $ 1.5\,Mpc $\times $ 0.1\,Mpc slices.  The dashed circles show $R_{\rm vir}$. {\it Bottom row:} gas particles shown in the same slice and colored by the logarithm of the ratio of turbulent energy to thermal energy, $E_{\rm turb}/E_{\rm therm}$. The total $E_{\rm turb}/E_{\rm therm}$ for dominant bubbles volume for each region is 12\% for  $\epsilon_{\rm 5, LBH}$, 18.8\% for $\epsilon_{\rm 15, EBH}$, and 23.8\% for $\epsilon_{\rm 50, EBH}$ (more details in the text). Note that bubbles are outlined and sometimes nested bubbles are displayed. The orientation of the main cosmological filament embedding the host DM halo and the expanding bubbles is indicated by the nearly horizontal white dashed lines. Black dashed lines show the bubbles opening angles. 
\label{fig:vort}}
\end{figure*}

Figure\,\ref{fig:cgm_props} shows evolution of the median thermal gas pressure, $P=nkT$, and entropy, $K=T/\rho^{2/3}$, in the CGM, normalized by $\epsilon_0$. We have performed a careful comparison between this Figure and 2-D projections of a 3-D expansion of the bubbles. We observe that bubbles can comprise a large fraction of the CGM volume at various times. So they can affect both the temperature and density within the CGM. While the temperature inside the bubbles is always substantially higher than that of the ambient medium, the density effect can be either in reduction of the average density or not, depending on whether a bubble is still propagating within the CGM, or the bubble boundaries lie outside the CGM. 

In this Figure, we can see that both $P$ and $K$ reflect the bubble presence, as they propagate into and across the CGM. We observe a largely different response from the EBH and LBH sequences, which is a reflection of the same trend present in Figure\,\ref{fig:outflow}. While the $\epsilon_{\rm 15,LBH}$ and $\epsilon_{\rm 50,LBH}$ models experience a sharp decline in thermal pressure after $z\sim 2$ with respect to $\epsilon_0$, all EBH models experience a sharp increase during this time, this is clearly related to evolution of the f$_{\rm CGM}$ in these models: the former shows a sharp decline, the latter a sharp increase in f$_{\rm CGM}$. Earlier, we have commented that this evolution affects $\dot M_{\rm out}$.  

Bubble expansion into the CGM is always associated with increasing gas temperature, more so in the EBH models. The density decrease in the LBH models is more substantial than in the EBH models --- this results in an overall decrease in the normalized $P$ there.  Smaller variation in density of the EBH models, results in the overall increase in the normalized thermal pressure. The spikes in thermal pressure are stronger where the bubble shock fronts traverse the dense CGM gas for the first time.   

The entropy in the lower frame of Figure\,\ref{fig:cgm_props} is indicative of the presence of a hot gas in the CGM, and thus is strongest where we have low density, hot gas residing inside the bubble which occupies the entire CGM volume and beyond. Associated increases in entropy tend to indicate the presence of new nested bubbles.

Evidently, the density and temperature, and thus the pressure and entropy, approach the $\epsilon_0$ model when no new bubbles have been triggered, e.g., between $z\sim 1-0.5$ in the $\epsilon_{\rm 15,LBH}$ model. This can be traced to a slightly earlier time when $\dot M_\bullet$ drops substantially due to a lack of available gas in the galactic centers (see Figures\,\ref{fig:BHmass} and \ref{fig:fgas}. From the animation linked in Figure\,\ref{fig:VRad}, we do not see the propagation of significant new bubbles during this time. This shows that the CGM can recover in a few Gyr to a state similar to that in $\epsilon_0$ through cooling and gas accretion. However, a relic of the past bubbles remains through a sustained increase in gas metallicity. 

To further investigate interaction between the inflating bubbles and the ambient gas, we invoke vorticity, $\vec\omega=\vec\nabla \times {\vec v}$, and show that there is enhanced turbulence along the bubble boundaries, triggered by the Kelvin-Helmholtz instability, where a hot interior bubble gas encounters dense, cooler exterior gas. Vorticity has been enhanced by over an order of magnitude, and clearly delineates the boundaries of bubbles, sometimes even displaying the nested bubbles. In Figure\,\ref{fig:vort} we show three 1\,Mpc $\times $ 1\,Mpc and 1.5\,Mpc $\times $ 1.5\,Mpc slices, 0.04--0.1\,Mpc deep, of vorticity maps for the CGM, centered on the main galaxy. The black color inside the bubbles indicates regions with a low density gas, compared to typical bubble gas in the CGM. On the CGM scale, and especially outside $R_{\rm vir}$, bubbles propagate orthogonally to the associated cosmological filament. Note that incoming filamentary gas experiences amplification of vorticity, in agreement with \citet{bi24}.

Finally, to estimate the overall importance of turbulence induced in the CGM, we have calculated the ratio of turbulent-to-thermal energy, $E_{\rm turb}/E_{\rm therm}$, in the CGM slices shown in  Figure\,\ref{fig:vort}. The turbulent energy calculation we based on the kernel, i.e., using the nearest 64 neighbors for each gas particle in the CGM. Next, the 3-D velocity dispersion $\sigma$ has been calculated, which automatically removes the bulk motion. This procedure was looped for all particles to obtain their turbulent velocities and energies, $E_{\rm turb}$. The ratio $E_{\rm turb}/E_{\rm therm}$, has been obtained first per gas particle, then integrated over the approximate bubble volume at the specific times shown in the upper frames of Figure\,\ref{fig:vort}.
 
For $\epsilon_{\rm 5,LBH}$ model, this ratio is equal to $\sim 12\%$, for $\epsilon_{\rm 15,EBH}$ it is 18.8\%, and for $\epsilon_{\rm 50,EBH}$ it is 23.8\%. Comparable $\epsilon_{\rm 0}$ models have $E_{\rm turb}/E_{\rm therm}\sim 8\%$, 18.3\% and 18.7\% at the same time and the same volume. Hence, while the thermal energy in the CGM always dominates, the turbulent energy is not negligible. The turbulence maps allow for observations to detect the current and possibly past AGN activity in the form of expanding bubbles.

\section{Discussion}
\label{sec:discuss}

We carried out a set of high-resolution cosmological zoom-in simulations to study the effects of AGN mechanical and thermal feedback in Seyfert galaxies with log\,$M_*/M_{\odot}\sim 11$ at $z=0$. Using identical initial conditions for all models, we have varied only the SMBH seeding time, and AGN and SN feedback efficiency. For comparison,  non-AGN models have been run (Table\,\ref{tab:models}).  The AGN jets have triggered expanding bubbles which cross the ISM, CGM, and beyond. While jets influenced mostly the inner galaxies, the inflating bubbles exerted a pronounced effect on the ISM and CGM.

We start this section by listing our main results and follow up with discussion. The AGN feedback effects on major scaling relations in modeled galaxies will be addressed separately (Goddard et al., in prep.). 

$\bullet $ The stellar mass of modeled galaxies at $z=0$ anti-correlates with the strength of the AGN feedback, i.e., for $\epsilon = 0.05-0.5$ it varies by a factor of 2--3, compared to non-AGN models. This result affects the galaxy stellar mass function (SMF) around the `knee,' at $M_*\sim 10^{11}\,M_\odot$, where the SMF falls steeply \citep[e.g.,][]{weaver23}, thus shifting the modeled mass bin to the left of the knee. The SFRs follow the same trend with $\epsilon$, although their decline after $z\sim 2$ can be reversed for limited times due to the ISM replenishing.

$\bullet $ Mass concentration of modeled AGN galaxies, based on $R_{1/2}$, anticorrelates with $\epsilon$. Moreover, the EBH models are less concentrated compared to the LBH --- thus a prolonged AGN stage, based on the BH seeding time, acts to inflate the galaxies. The SN feedback strength adds slightly more to this trend.  The surface density of the stellar disks averaged over the $z = 2-0$ period displays a substantial deficiency at smaller radii compared to the non-AGN models, reflecting a secular damping of the disk SFR at smaller radii, $\ltorder 10$\,kpc, due to the AGN feedback. Although the SFR in the bulge has essentially switched off after $z\sim 2$, the disk component experiences a more gradual decline in SFR until $z\sim 0.5$. 

$\bullet $ The SMBHs final masses, $M_\bullet\sim 10^7 - 10^8\,M_\odot$, correlate directly with $\epsilon$, and this trend is more profound for the early seeded BHs. Dispersion of SMBH final masses is also larger for the EBH sequence compared to the LBH one --- a direct consequence of EBH seeds evolving over a longer time period. The most massive SMBH and the least massive one is from the EBH sequence.

$\bullet $ Seyfert jets have a dual effect on the ISM, namely, by a direct interaction with the gas and by triggering expanding bubbles. The jets are mostly contained within the galaxies or so, but the bubbles driven by the energy deposited by the jets propagate across the CGM and beyond, expelling some of the gas from the galaxy (Figs.\,\ref{fig:fgas} and \ref{fig:VRad}) and slowing down the SFR (Fig.\,\ref{fig:SFR}). These shocks also interfere with cosmological accretion flows. 

$\bullet $ Bubbles expand mostly perpendicularly to the cosmological filaments outside $R_{\rm vir}$, generating strong turbulence within the CGM --- a result of the Kelvin-Helmholtz instability at the two-fluid interface, with $E_{\rm turb}/E_{\rm therm}\sim 10-25\%$. Dynamically, bubbles reduce the gas density in the CGM, driving outflows across $R_{\rm{vir}}$. Thermodynamically, they increase the CGM entropy, leading to an increase or decrease of thermal pressure. Furthermore, bubbles increase the CGM metallicity up to a factor of $2-3$ compared to non-AGN models. Seeding the SMBH earlier and increasing the SN feedback make a significant impact on the CGM metallicity.   

Next, we explore certain quandaries raised by our results and put our simulated galaxies into context with observed galaxies. 

All the stellar galaxy masses of AGN models converge at low $z$ (Fig.\,\ref{fig:starmass}). Among themselves, they differ less than with respect to non-AGN models (with exception of $\epsilon_{\rm 5,LBH}$), despite a sharp increase in the AGN feedback by an order of magnitude. Thus, while stronger feedback has a more dramatic effect on the ISM, it operates over shorter time periods. 

AGN jets and bubbles deposit their energy and momentum on a wide range of scales in the gas, changing its distribution and properties, hence affecting the distribution of forming stars, and galactic morphology. As shown in Figure\,\ref{fig:BHmass}, most of the energy from the jet particles is deposited in the central few kpc. This deposition reduces the gas and H$_2$ concentration, and thus also the SFR in the central few kpc.  In the most catastrophic cases of gas loss, such as in both $\epsilon_{50}$ models at $z\sim 2$, the gas abundance in the central regions recovers on different timescales, from $\sim 1$\,Gyr to 9\,Gyr. 

The SF resumes at large radii first, due to an excess of gas there (Fig.\,\ref{fig:SurfDen}). Jet extent, $R_{\rm jet}$, anti-correlates with the average density within $R_{\rm jet}$ (Fig.\,\ref{fig:Rjet_ncgm}). Some of the gas is pushed out of halos, increasing the outflow rate and decreasing the net inflow rate across $R_{\rm vir}$ --- interaction between the bubbles and the CGM gas has been shown in Figure\,\ref{fig:vort}, where boundaries of the bubbles are clearly delineated by increased vorticity.

The stellar mass concentration measured by $R_{1/2}$ was found to be sensitive to the parent DM halo spin and its merger history \citep[e.g.,][]{mo98,naab09}. Having this in mind, we run our models from identical cosmological conditions to remove these effects. Therefore, we attribute the correlation between $R_{1/2}$ and $\epsilon$ to the energy and momentum deposited by the jet and the associated bubbles (Fig.\,\ref{fig:galR}). Because the merger history is basically identical in all our models, the observed peak in $R_{\rm 1/2}$ with respect to $\epsilon_0$ models around $z\sim 2$ is triggered by the increased jet and bubble activity. 

The by-product of the ISM interaction with the jets is a non-monotonic evolution of gaseous disk sizes, as seen in Figure\,\ref{fig:galaxies}. The most intense feedback around $z\sim 2$ is associated with the smallest extent of the gaseous disks. Moreover, at these redshifts, the median entropy of the CGM in the AGN models is larger up to a factor of $\sim 30$ over that in the non-AGN models.


Now we turn to evolution of the three $\epsilon_5$ models, including the test one. The $\epsilon_{\rm 5,LBH}$ model most closely resembles the non-SMBH evolution, while $\epsilon_{\rm 5,EBH}$ follows the other AGN models (Figs.\,\ref{fig:starmass},\ref{fig:SFR},\ref{fig:fgas},\ref{fig:Mh2},\ref{fig:MassKD},\ref{fig:vlos}). What is the reason for this behavior?    

To answer this question, we note that $\epsilon_{\rm 5,LBH}$ is the only AGN model which does not form bubbles until $z\sim 0.2$ (Fig.\,\ref{fig:VRad}). We ran a test model, $\epsilon_{\rm 5,EBH}$(test), which has the same AGN and SN feedback as $\epsilon_{\rm 5,LBH}$, with only difference in the BH seeding time. Yet, the test model develops bubbles shortly after $z\sim 6$, and evolves similarly to $\epsilon_{\rm 5,EBH}$ in all respects. Hence, only the seeding time of the BHs in $\epsilon_{\rm 5,LBH}$ and $\epsilon_{\rm 5,EBH}$, and associated events, i.e., interaction between the ISM and jets, separate the models.

Indeed, $L_{\rm jet}(t)$ differ between $\epsilon_{\rm 5,LBH}$ and $\epsilon_{\rm 5,EBH}$ (Fig.\,\ref{fig:BHmass}). Analyzing the AGN feedback in $\epsilon_{\rm 5,LBH}$ reveals a very different evolution of the SMBH spin axis angle, both polar and azimuthal, which determine the direction of the jet particle injection, and, therefore, lead to different jet-ISM interaction. Evolution of the normalized thermal pressure and entropy in the ISM (not shown here) supports this view --- this can even be traced to the CGM (Fig.\,\ref{fig:cgm_props}).


Note that identical initial cosmological conditions lead to a different evolution in $\epsilon_{\rm 50,LBH}$ and $\epsilon_{\rm 50,EBH}$ models which exhibit a history of frequent powerful expanding bubble formation. Their evolution, and especially the final gas distribution, are quite different (Figs.\,\ref{fig:VRad} and \ref{fig:galaxies}). Both galaxies experience a catastrophic gas loss (Fig.\,\ref{fig:fgas}) around $z\sim 2.5 - 1.5$, when strong bursts of SN and jet activity coincide with interactions (Fig.\,\ref{fig:BHmass}). However, $\epsilon_{\rm 50,EBH}$ has recovered the gas by $z\sim 1$, while in $\epsilon_{\rm 50,LBH}$, the gas has not returned fully even by $z=0$ (Figs.\,\ref{fig:Mh2} and \ref{fig:galaxies}). What is the reason for such a diverging evolution? 

From the above comparison of $\epsilon_5$ models, having identical initial conditions is not a recipe for similar evolution over the Hubble time. For the $\epsilon_{50}$ models we point to Figures\,\ref{fig:BHmass} and \ref{fig:outflow}. In the former Figure, the $\epsilon_{\rm 50,EBH}$ jet particles breakout of the galaxy and propagate deep into the CGM, while $\epsilon_{\rm 50,LBH}$ jet is trapped within the galaxy. So, the energy deposition region differs profoundly. Following the ISM expulsion from these galaxies around $z\sim 3$, in the early-seeded model the gas remains outside the galaxy, in the innermost CGM. However, for the late-seeded model, the gas is pushed deep into the CGM and increases the {\it outflow} rate across $R_{\rm vir}$, which leads also to a decrease of gas fraction in the CGM by $\sim 60\%$ (Fig.\,\ref{fig:outflow}).  As a result, the gas is capable of returning and repopulating the galactic disk in $\epsilon_{\rm 50,EBH}$ by $z\sim 1$, but not in $\epsilon_{\rm 50,LBH}$. This also manifests itself in resumption of the SMBH growth, leading to a sharp increase in $\dot M_{\bullet}$, and consequently an increase of $L_{\rm jet}$ at $z\ltorder 0.5$, not seen in $\epsilon_{\rm 50,LBH}$.  Increase in $L_{\rm jet}$ creates powerful long-range jets and a number of nested bubbles that reduce the cold accretion from the cosmological filaments, affecting stellar disk size. Hence, minor evolutionary changes can lead to substantial changes in the final morphology of Seyferts, as well as to dispersion of the SMBH mass values at $z=0$.  

\subsection{Simulations vs observations of jetted Seyferts} 
\label{sec:compare}

Here we briefly compare our simulations with observations of jetted Seyferts in the local and early universe. Our SMBH seeds evolve from $10^6\,M_\odot$ by about 2 orders of magnitude during $z\sim 9-0$, with dispersion in $M_\bullet$ by a factor of ten.  This is consistent with the range of SMBH masses in observed Seyferts \citep[e.g.,][]{Foschini2020,varglund22}. Initial $M_\bullet$ has been motivated by direct collapse models of SMBH formation \citep[e.g.,][]{begelman09,begelman10,choi13,inayoshi20}. During the simulation time, $L_{\rm jet}$ and f$_{\rm Edd}$ vary over the range of $10^{40}-10^{42}\,{\rm erg\,s^{-1}}$ and $0.3 - 10^{-4}$, respectively. This also trends with both luminous, $\dot M_\bullet\sim 10^{-1-2}\,M_\odot\,{\rm yr^{-1}}$, and low-luminosity, $\dot M_\bullet\sim 10^{-3}\,M_\odot\,{\rm yr^{-1}}$, Seyferts which are characterized by $L_{\rm jet}\ltorder 10^{43}\,{\rm erg\,s^{-1}}$.  The jet angles vary with time and show no particular preference with respect to the galaxy disk, as predicted \citep[e.g.,][]{osterbrock82,kinney00,schmitt02}.  
 
We observe that most of our galaxies maintain disk-dominated morphologies for much of their evolution (Fig.\,\ref{fig:vlos}). Only the strongest feedback, $\epsilon_{\rm 50}$ models, trend toward a spheroidal morphology at late times, $z\ltorder 1.5$. The majority of observed jetted Seyferts are likewise hosted by disk-like galaxies, with spiral structure \citep{varglund22}, though  there are also examples of elliptical hosts \citep{Foschini2020}.  Once a statistically significant number of host morphologies is gathered, it should be tested for any correlation between AGN jet power and morphology, as observed in this work --- this refers to major scatling relations, including the SMF.  However, the instantaneous jet power is not a good indicator of the total injected energy over time.  Instead, one could look for indicators of previous sustained strong bubbles, such as radio relics or a low-density CGM. As SMBHs are hosted by all galaxies around the SMF knee, the Seyfert can be an active phase of otherwise `normal' galaxies.

\citet{olguin-iglesias20} suggests that jet detections via `radio-loudness' tend to come from jets during secular evolution, while minor-merger triggered jet activity tends to be $\gamma$-ray detected. While our strong feedback effects occur during high $z$, including periods of frequent interactions, the largest-scale, most powerful jets occur at $z < 0.5$, in the absence of significant interactions, under a more quiescent, secular evolution.  

It is argued that jets in Seyferts may be triggered by interactions, 
\citep{järvelä18b}. A tendency toward central dust reddening in IR observations can indicate recent minor mergers \citep{varglund22}. However, the presence of stellar bars would lead to similar effects \citep[e.g.,][]{knapen95,knapen00,laine02}. In all these cases, the gas falls in due to gravitational torques. 

We find reduced central $\Sigma_{\rm gas}$ and $\Sigma_*$ in the presence of AGN feedback, and the degree of relative reduction scales with $\epsilon$. Likewise, we find periodic reductions in the overall H$_2$ mass of our galaxies due to the jet feedback. Depletion of H$_2$ from the central 0.5\,kpc by a jet has been observed in Seyfert\,2 galaxy IC 5063 \citep{morganti15}, and in the central 6\,kpc of NGC\,2639 \citep{rao23}. Various additional accounts exist of interactions between jets and the gaseous disk of their host \citep[][]{venturi21,Nesvadba2021,Murthy2022,girdhar22,audibert23}. 

The SFR in our models peaks between $z\sim 5-2$, and stays in the range $0.001-30\,M_{\odot}\,{\rm yr^{-1}}$, covering the full range from starburst to quenched SF. We find a broad spectrum of overall effects to SF in our AGN models.  For the $\epsilon_{\rm 5,LBH}$ model, the SF is largely unaffected by the presence of the jet feedback.  Whereas for the $\epsilon_{\rm 50,LBH}$, the galaxy remains fully quenched for $\sim 9$\,Gyr. Other AGN models lie between these two extremes, but all display the strongest reduction to SF near the center of the galaxy. Similarly, observations have found a broad range of effects to SF from jet feedback. In some cases, SFR seems boosted or unaffected by the jet presence \citep[e.g.,][]{caccianiga15,berton20,morganti21,joseph22,Kurian2024}. Others observe that SFR is reduced or even quenched, particularly in galaxies with log\,$M_*/M_\odot\,> 11$ \citep[e.g.,][]{Nesvadba2021,rao23,jin25}.  

Our suite of simulations find that bubbles from AGN jets affect the CGM by enhancing the gas metallicity, injecting entropy, enhancing or reducing the thermal pressure, and driving shocks. We also see enhanced vorticity along the bubble edges (Fig.\,\ref{fig:vort}). Many observed jetted Seyferts show the presence of extended radio relics, bow shocks, and dense gas distributed along the jet direction \citep[e.g.][]{cecil00,rosario10,xanthopoulos10,biny19,zeng23,Duggal2024,marques24}, showing that these small-scale jets do launch large-scale bubbles and expel cold gas from the galaxy into the CGM. Numerical studies have also shown that cocoons can enrich gas in the CGM through the expulsion of high-metallicity disk gas \citep[e.g.,][]{choi20,appleby21}. This mechanism can be responsible for the observed IGM metallicities which exceed those of simulations that exclude such feedback. There is a need for high-resolution multi-wavelength observational studies of the local environments around jetted galaxies, so theoretical predictions can be further substantiated.

There is a lack of observational studies involving the CGM properties around AGN jet-hosting galaxies. But theoretical and numerical works on the influence of the bubbles on the galaxy environment exist, e.g., supported by Hitomi X-ray telescope detection of AGN feedback-generated turbulence of $\sim 4\%$ of the {\it total} pressure in Perseus cluster \citep{hitomi16}, and XRISM observations of Hydra\,A \citep{rose25}, in agreement with our results. Simulated jet-ambient gas interactions confirm analytic predictions of vortex formation where strong mixing occurs \citep{antonuccio10b}, as we also observe in this work.  Observationally, this could be detected through line-of-sight measurements of velocity fields, where enhanced turbulence can alter the spectral line profiles.  Similarly, radio observations can show enhancements in brightness, where higher vorticity leads to increased synchrotron emission \citep[e.g.,][]{borse21}.  

We find that our large-scale galaxy morphologies and AGN properties closely reflect observations of local jetted Seyferts. As such, they should make valid predictions for early universe evolution of Seyferts. It has been suggested that jetted NLS1s are young AGN \citep[e.g.,][]{Foschini2020, olguin-iglesias20,jarvela21}.  However, our simulations show that these galaxies maintain the properties of a typical Seyfert for many Gyrs, and also typically peak in their luminosity around $z\sim 5-2$, as do quasars. As such, we expect to find jetted Seyferts at high redshifts, although detection of these low-luminosity sources is tricky, especially when the jet is obscured.  A pool of high-$z$ Seyfert observations grows with JWST \citep[e.g.,][and refs. therein]{lin25}. Future observations with advanced radio telescopes, e.g., Square-kilometer Array - SKA \citep{berton2016}, and faint $\gamma$-ray source detections, using Cherenkov Telescope Array - CTA, \citet{romano20}, will greatly enhance the number of known jetted Seyferts and thus our ability to compare theory with observations. 

Finally, we comment about the duty cycle of jets. Current understanding is that the innermost accretion disk accelerates and collimates the outflow due to the magnetic fields \citep[e.g.,][]{shakura73,blandford82,contopoulos95,fabian12}. This encompasses the whole range of AGN, from quasars to microquasars \citep[e.g.,][]{livio03}, possibly including the direct collapse pre-SMBH objects \citep{luo24}. However, under what conditions are such jets triggered? In our modeling, we take the simplistic approach that jet luminosity scales with the SMBH accretion rate, and ignore the possible modifications in the physical nature of these jets. Hence, our jets operate as long as the SMBH accretes, ignoring such `nuances' as e.g., radio-mode. 

To summarize, there has been significant previous numerical work on the effects of AGN jets in galaxies \citep[e.g][]{mukherjee18,su21,talbot22,irodotou22,qutob23,wellons23,su24,byrne24}. These have found strong effects to the host galaxy and environment on short timescales, or at $z=0$ \citep[e,g.,][and refs. therein]{goddard25}. The current work emphasizes the long-term evolution of the thermodynamic and morphological properties of Seyferts and their environment during $z\sim 9-0$, considering the jet-driven expanding bubbles. We show that many of the effects to the galaxy and CGM are transient, such as influences to the SFR and thermodynamic changes to the CGM. These can recover to some degree, given a significant period of reduced activity. However, some changes are cumulative with time, and scale with either AGN jet deposited energy, SMBH seeding time, or both, such as $\Sigma_*$, $\Sigma_{\rm gas}$, M$_*$, CGM metallicity, etc.  As such, it is possible these can be used as tracers for long-term activity in observed jetted Seyferts when compared against the greater galaxy population.  

\begin{acknowledgments}
We thank Phil Hopkins for providing us with the latest version of the code, and are grateful to Alessandro Lupi, Da Bi, Kung-Yi Su, Paul Torrey, Xingchen Li, and Clayton Heller for their help with GIZMO. E.R.D. acknowledges support of the Collaborative Research Center 1601 (SFB 1601 sub-project C5), funded by the Deutsche Forschungsgemeinschaft (DFG, German Research Foundation) – 500700252. This work used Expanse CPU at San Diego Supercomputer Center (SDSC) through allocation PHY230135 from the ACCESS program, which is supported by NSF grants 2138259, 2138286, 2138307, 2137603, and 2138296 \citep{boerner23}, and by the University of Kentucky Morgan Computing Cluster. We are grateful for help by Vikram Gazula at the Center for Computational Studies of the University of Kentucky.
\end{acknowledgments}  

\section*{Data Availability}
 The data used for this paper will be shared upon any reasonable request which intends the data to be used for scientific endeavors. Requests can be made by email to the authors. The available data includes figures, ascii tables of galaxy and CGM data, and simulation snapshot files. The method of data transfer will be determined on a case by case basis as we will need to determine an appropriate method based on the specific request.

\bibliography{references}{}
\bibliographystyle{aasjournal}


\end{document}